\documentclass[aps,pra,reprint]{revtex4-2}
\usepackage{graphicx} 
\usepackage{amsmath}
\usepackage[normalem]{ulem}
\usepackage{comment}
\usepackage{csquotes}
\usepackage{multirow}
\usepackage{xcolor}
\usepackage{setspace}
\usepackage{hyperref}
\usepackage[margin=1in]{geometry}
\usepackage[normalem]{ulem}
\def\Ln{\Delta L_n}
\def\La{\Delta L_a}
\def\Ls{\Delta L_s}
\def\ea{\eta_{\rm ani}}
\def\half{\frac{1}{2}}

\newcommand{\braket}[2]{ \langle #1 | #2 \rangle} 

\usepackage{xcolor}

\begin{document}

\title{Optical microcavity characterization via resonance spectra and modes}

\author{Jonah Post$^1$}
\author{Chunjiang He$^1$}
\author{Corné Koks$^{1,3}$}
\author{Rudi van Velzen$^1$}
\author{Andrea Corazza$^2$}
\author{Yannik L. Fontana $^2$}
\author{Marcel Erbe$^2$}
\author{Richard J. Warburton$^2$}
\author{Martin P. van Exter$^1$}
\affiliation{1. Huygens-Kamerlingh Onnes Laboratory, Leiden University, P.O. Box 9504, 2300 RA Leiden, The Netherlands}
\affiliation{2. Department of Physics, Basel University, Klingelbergstrasse 82, 4056 Basel, Schweiz}
\affiliation{3. Department of Electrical and Photonics Engineering, Technical University of Denmark, Ørsteds Plads 345A, DK-2800 Kgs. Lyngby, Denmark}

\date{\today}

\begin{abstract}
This paper describes how resonance spectra and mode profiles can be used to characterize and quantify the mode-shaping effects in open-access plano-concave optical microcavities. 
The presented semi-analytic theory is based on the application of perturbation theory to the roundtrip evolution of the optical field. 
It includes various mirror-shape and nonparaxial effects and extends the nonparaxial theory presented in ref. \cite{VanExter2022} and verified in ref. \cite{Koks2022b} to the common case of an anisotropic Gaussian mirror. 
The presented measurements and analyses of resonance spectra and mode profiles demonstrate how the different mode-shaping effects can be individually distinguished and quantified.  
Spin-orbit coupling, which is one of the nonparaxial effects, is prominently visible in the intriguing polarization patterns of the resonant modes, while polarization tomography yields the shape-induced birefringence and associated polarization splitting of the fundamental modes. 
\end{abstract}

\maketitle

\section{Introduction}
\label{sec:introduction}

Optical microcavities can resonantly trap light and thereby enhance the light-matter interaction of intra-cavity emitters~\cite{Vallance2016,Flatten2018,Najer2019,Wang2019,Vogl2019, Haussler2021} and increase the collection efficiency in the resonant cavity mode(s)~\cite{Ruf2021}.
This makes them essential for experiments in quantum optics and quantum information \cite{Vahala2003,Tomm2021,Favero2009,Stapfner2013,Flower2012}. 

The simplest tuneable microcavity is the open-access plano-concave Fabry-Perot microcavity, with a flat and a concave mirror (radius of curvature $R_m$), both highly reflecting, spaced at distance $L$ with $\lambda < L < R_m$.
The plano-concave geometry is relatively insensitive to alignment, while the open access allows easy length tuning and addition of intra-cavity emitters. 
The technical challenge of open cavities being relatively sensitive to mechanical vibrations is gradually overcome and cavity stabilities below 10 pm have been demonstrated \cite{Najer2019,Fontana2021,Pallmann2023,Fisicaro2024}.

High-finesse optical Fabry-Perot cavities have become available after the development of super-polished mirrors.
The first experiments were performed with alkali atoms in relatively large cavities \cite{Raizen1989,Zhu1990,Hood2001}.
Since then, the mirror radii and cavity lengths have shrunk to micrometer size and cavity mode-shaping effects have become more prominently visible.
These effects are discussed below.
The presented theory should also be valid for open Fabry-Perot cavities with emitters grown on top of one of the mirrors, like the open semiconductor cavities in ref. \cite{Greuter2014,Barbour2011}.
It could even work for composite cavities with micrometer-thin diamond slabs \cite{Ruf2021,Flagan2022}, although reflections at the diamond-air interface are then likely to codetermine the cavity modes and their resonance conditions. 


The interpretation of resonance spectra and modes can be performed in three levels of complexity.
The simplest interpretation uses the paraxial description for a cavity with anisotropic mirrors. 
This analytic description predicts resonance spectra with Hermite-Gaussian HG$_{mn}$ eigen modes that are equidistantly spaced within each transverse $N=m+n$ group and frequency degeneracy for spherical mirrors. 
The next level of complexity introduces arbitrary mirror shapes but retains the paraxial propagation. 
This general analysis, which was introduced by Kleckner et al. \cite{Kleckner2010}, can only be done numerically.
It has o.a. been used to model the observed resonances and mode profiles in aligned \cite{Greuter2014,Benedikter2015,Mader2015,Mader2022} and misaligned FP microcavities \cite{Hughes2024}.
The paraxial description works well for cavities with $R_m \gg \lambda$ but is incomplete for cavities with very small $R_m$ because it neglects nonparaxial effects and uses scalar fields. 
This naturally brings us to the third level of complexity, a level that includes nonparaxial effects. 
This level is needed to properly interpret the spectral fine structure within each $N$ group of modes, to understand why these modes are typically not equidistant and spectrally split even in cavities with perfect spherical mirrors, and to understand the intriguing polarization patterns of the eigen modes.
If any of these effects show up, nonparaxial effects might be visible in your microcavity. 
At this final level of complexity, all nonparaxial and some mirror-shape effects can be treated semi-analytically by applying perturbation theory to the well-known paraxial solutions \cite{VanExter2022}.   

The importance of nonparaxial effects was demonstrated experimentally by Koks et al. \cite{Koks2022b}, who studied mode formation in short cavities with close-to-spherical mirrors with small $R_m$.
In their cavities, the nonparaxial effects dominated over the mirror-shape effects and the observed spectral fine structures and mode profiles resembled the theoretical predictions \cite{VanExter2022}.

The relative strength of nonparaxial effects in microcavities can be estimated with the following guideline.
The splitting $\Delta L_{\rm trans}$ between consecutive groups of transverse modes yields the dimensionless ratio $\eta_{\rm par} = \Delta L_{\rm trans}/L$ for a length scan, or $\eta_{\rm par} = \Delta f_{\rm trans}/f$ for a frequency/wavelength scan.  
This $\eta_{\rm par} \approx 1/k\sqrt{LR_m}$, with $k=2\pi/\lambda$, measures the paraxiality or mean square opening angle of the fundamental mode.
The nonparaxial correction is order $\Delta L_{\rm non}/L \propto \eta^2_{\rm par}$, yielding a relative correction order $\eta_{\rm par}$ on the transverse-mode splitting.
This correction should thus become visible in resonance spectra when $F \lambda/R_m > 10$, where $F$ is the modal finesse. 
However, the nonparaxial effect might still be difficult to spot if it is overwhelmed by other mode-shaping effects, like the anisotropic splittings that occur in cavities with astigmatic mirrors.

This paper focuses on the regime where mirror-shape and nonparaxial effects have comparable magnitudes, making them challenging to disentangle. 
It aims to provide a practical framework for experimental physicists to address these complexities and quantify the contributing effects.
It presents recipes and equations to characterize the precise geometry, including the distance between the mirrors, the radius of curvature of the concave mirror, and its anisotropy and spherical aberration. 
This characterization is based on resonance spectra measured by monitoring the transmission $T(L)$ while scanning the cavity length $L$ at fixed $\lambda$, although wavelength tuning would also work.
It is supplemented by the analysis of spatial profiles of the transmitted resonant modes measured with a polarization-resolving camera. 

This paper is structured as follows.
Section \ref{sec:theory} revisits the theory outlined in Ref. \cite{VanExter2022}, extends it with an aspheric correction that is crucial for the commonly-used smooth Gaussian mirror made by laser ablation, and emphasizes the role of symmetry. 
Readers primarily interested in the experimental results may skip directly to Sec. \ref{sec:setup}, which describes the experimental setup. 
Section \ref{sec:spectra} presents the measured resonance spectra (transmission and reflection), while Sec. \ref{sec:profiles} focuses on the polarization-resolved mode profiles. 
Section \ref{sec:comparison} compares the deduced mirror shape with the measured mirror shape, while Sec. \ref{sec:summary} summarizes the key findings and conclusions.

\section{Theory}
\label{sec:theory}

\subsection{Paraxial description with spherical mirror}
\label{sec:paraxial}

For a plano-concave cavity with a spherical mirror, the paraxial description of the round-trip evolution yields solutions in the form of Hermite-Gaussian (HG) or Laguerre-Gaussian (LG) modes.
These solutions are labeled by their longitudinal mode number $q$ and their transverse order $N$, where $N = m+n$ for HG$_{mn}$ modes and $N=2p+\ell$ for LG$_{p\ell}$ modes \cite{Siegman}. 
The boundary conditions at the mirrors determine the Rayleigh range $z_0 = {\scriptstyle \frac{1}{2}} kw_0^2 = \sqrt{L(R_m-L)}$, the waist $w_0$ at the plane mirror, the beam width $w(z) = w_0 \sqrt{R_m/(R_m-L)}$ at the curved mirror, and the matched wavefront radius $R(z) = z+z_0^2/z = R_m$ at $z=L$.

For these cavities, the paraxial resonant cavity length of each $j = (q,N)$ mode is
\begin{equation}
\label{eq:concave-resonance}
    L(q,N) = \left[ q + (N+1) \frac{\chi_0}{\pi} \right] \frac{\lambda}{2} \,,
\end{equation}
where $\chi_0 = \chi_0(L) =\arcsin{\sqrt{L/R_m}}$ is the single-pass Gouy phase lag of the fundamental mode. 
The second term describes the paraxial contribution 
to the resonance length.  
It's relative strength 
\begin{equation}
    \eta_{\rm par} = \frac{\Delta L_{\rm par}}{L} = \frac{1}{kL} \arcsin{\sqrt{\frac{L}{R_m}}} \approx \frac{1}{k\sqrt{LR_m}} \,, 
\end{equation}
where $\Delta L_{\rm par} = \chi_0/k$ is linked to the mean-square opening angle of the fundamental mode 
\begin{equation}
\label{eq:theta^2}
    \langle \theta^2 \rangle_{N=0} = \frac{1}{k\sqrt{L(R_m-L)}} \approx \frac{1}{k\sqrt{LR_m}} \,, 
\end{equation}
and to the mean-square radius of that mode on the flat mirror
\begin{equation}
\label{eq:r^2}
    \langle r^2 \rangle_{N=0} = \gamma_0^2 = \frac{z_0}{k} = \frac{\sqrt{L(R_m-L)}}{k} \,, 
\end{equation}
with $\gamma_0 = w_0/\sqrt{2}$ and intensity profile $I(r) = I(0)\exp(-r^2/\gamma_0^2)$.
The dimensionless parameter $\eta_{\rm par}$ quantifies the ``nonparaxiality" of the mode.
The mean-square opening angle and mean-square radius of the higher-order modes scale with $(N+1)$.

In a typical experiment, one measures the cavity transmission while scanning the cavity length. 
One thus obtains various relative resonant cavity lengths $L_j \approx L(q,N)$, but does not yet know the absolute $q$ nor the offset length $L_{\rm off}$. 
This offset length can be determined from the $q$-dependence of the transverse-mode splitting 
\begin{eqnarray}
\label{eq:Ltrans}
     \Delta L_{\rm trans} & \equiv & L(q,N+1)-L(q,N) \nonumber \\
     & \approx & \Delta L_{\rm par} \approx \frac{\chi_0(L)}{k} \,.   
\end{eqnarray}
The approximate signs indicate that the second line in this equation uses the paraxial approximation and neglects a small second-order correction; a correction that also makes the free spectral range of the cavity a tiny bit larger than $\lambda/2$.

Appendix \ref{sec:appendix-Calibration} describes how raw transmission data can be transformed into $T(L)$ spectra, using the observed transmission peaks as ruler. 
This procedure corrects for the nonlinearity of the piezo scan and the optical penetration in the Distributed Bragg Reflectors (DBRs).
The latter correction is subtle because the optical penetration depth depends on the property of interest \cite{Koks2021}.
The resonant length $L(q,N)$ in Eq. (\ref{eq:concave-resonance}), for example, is actually the distance between two nodes of the optical field and hence differs from the surface-to-surface distance between the mirrors by the sum of their phase penetration depth $L_\varphi$. 
However, the Gouy phase $\chi_0(L)$ depends on the position of the modal waist in the DBR and, hence, on the sum of their modal penetration depths $L_D$. 
The combination of these effects yields
\begin{equation}
\label{eq:Gouy2}
    \sin^2{[\chi_0(L)]} = \frac{L(q,N)+L_{\rm pen}}{R_m} \,, 
\end{equation}
where $L_{\rm pen} = (L_{D 1}+L_{D 2}) - (L_{\varphi 1} + L_{\varphi 2})$. 
Appendix \ref{sec:appendix-Calibration} discusses this equation in more detail.
It shows how a measurement of $\chi_0(L)$ versus $L$ and a comparison with Eq. (\ref{eq:Gouy2}) allows one to determine the radius of curvature $R_m$ from the slope and the offset $L_{\rm off}$ (and modal $q$'s) from the axis crossing. 
It also shows that the ``combined effective penetration depth" $L_{\rm pen} \approx 0.38 \, \lambda/2$ is relatively small for our DBRs (see Appendix \ref{sec:appendix-Calibration}). 

\subsection{Mirror anisotropy and spherical aberration}
\label{sec:mirror-shape}

Equation (\ref{eq:concave-resonance}) is only an approximate description of the resonance condition, because it is based the paraxial approximation of optical propagation in a cavity with a perfect spherical mirror. 
Inspired by ref. \cite{VanExter2022}, we write the actual resonant length as 
\begin{equation}
     L_j =  L(q,N) + \Delta L_{j,{\rm fine}} \,, 
\end{equation}
where the fine structure 
\begin{equation}
\label{eq:L-fine}
        \Delta L_{\rm fine} = \Delta L_{\rm ani} + \Delta L_{\rm asp} + \Delta L_{\rm non} 
 ( + \Delta L_{\rm rest} ) \,,  
\end{equation}
describes how modes with the same $(q,N)$ values can still have different resonance lengths.
The contributions to the fine structure can be divided in two categories: (i) mirror-shape effects and (ii) nonparaxial effects. 
The mirror-shape effects, labeled as $\Delta L_{\rm ani}$ (anisotropy = astigmatism), $\Delta L_{\rm asp}$ (spherical aberration) and $\Delta L_{\rm rest}$ (rest), will be discussed in this subsection.
The nonparaxial effects $\Delta L_{\rm non}$ will be discussed in the next subsection.

The two most prominent mirror deformations are anisotropy and spherical aberration.
By choosing the $x,y$ axes along the astigmatic axes, we can model the shape of the common anisotropic Gaussian mirror as
\begin{eqnarray}
\label{eq:z}
    z_m(x,y) & = & h - h \exp[-(\frac{x^2}{2hR_x} + \frac{y^2}{2hR_y})] \nonumber \\
    & \approx & \frac{r^2}{2R_m} + \ea \frac{x^2-y^2}{2R_m} - \frac{r^4}{8hR_m^2} \,,
\end{eqnarray}
with average mirror radius $R_m = (R_x+R_y)/2$, anisotropy $\ea = (R_y - R_x)/2R_m$, with $|\ea| \ll 1$, and mirror depth $h$. 
Note that the $r^4$ term in Eq.~(\ref{eq:z}), which quantifies the amount of spherical aberration, is much larger than the $r^4/8R_m^3$ term in a similar equation for the spherical mirror.
Also note that higher-order Taylor terms are needed if the modes become too wide, i.e. for high $N$-modes in long cavities.
The fourth-order Taylor expansion only works well for $r^2 \ll hR_m$ and is disappointingly bad for larger $r$ (see inset in Fig. \ref{fig:Gouy_phase}).
In general, the parameter $h$ merely serves as a way to quantify the spherical aberration and should better be denoted as $h_4$ to distinguish it from the true mirror depth $h$.
This description then also works for mirrors with non-Gaussian shape.


Mirror anisotropy can be easily included in the paraxial description.
In the absence of other mode-shaping effects, anisotropic cavities support elliptical HG$_{mn}$ modes with waists $w_x \neq w_y$ and associated Gouy phases $\chi_x \neq \chi_y$.
This changes the $(N+1)\chi_0$ contribution to the resonance length into $(m+\half) \chi_x + (n+\half) \chi_y$, with $\chi_{x,y} = \arcsin{\sqrt{L/R_{x,y}}}$. 
We quantify the amount of anisotropy with $\ea$, as defined above via $R_x = R_m (1 - \ea)$ and $R_y = R_m (1 + \ea)$.
For $| \ea| \ll 1$, a Taylor expansion of $\chi_{x,y} = \arcsin{\sqrt{L/R_{x,y}}}$ yields 
\begin{equation}
\label{eq:DLani}
    \Delta L_{\rm ani} = (m-n) \Delta L_a \,\, ; \,\, \Delta L_a = \ea \frac{\tan(\chi_0)}{2k} \,.
\end{equation}
The mirror anisotropy will thus split the $(N+1)$ scalar HG$_{mn}$ modes within each $(q,N)$ group in a series of equidistant resonances spaced by $\Delta L_{\rm HG} = 2 \La \approx \ea \Delta L_{\rm par}$, because $(m-n)$ changes in steps of 2 and $\tan(\chi_0) \approx \chi_0$ for short ($L \ll R_m)$ cavities.
Modes with larger $m$ have larger resonance length $L_j$ when $\ea >0$, i.e. when $R_y > R_x$ such that $\chi_x > \chi_y$.
The spacing will not be equidistant though when other aberrations play a role. 
The parameter  
\begin{equation}
\label{eq:X}
    X(L) \equiv \frac{4\La}{\Ln} \approx 8 \pi \eta_{\rm ani} \left( \frac{\sqrt{L(R_m-L)}}{\lambda}\right)\,,
\end{equation}
compares the anisotropic splitting with the nonparaxial splitting $\Delta L_n = 1/2k^2R_m$ introduced below. 
The factor 4 has historic reasons and helps to remove factors 1/4 in some matrices; see below.

The Gaussian mirror shape creates a spherical aberrations that will modify the resonance lengths and deform the eigen modes towards Laguerre-Gaussian (LG) modes. 
The change in resonance length can be calculated with the generic equation $\Delta L = \langle \psi | \Delta z | \psi \rangle$ presented in ref. \cite{VanExter2022}, where $\Delta z$ is the mismatch between the mirror shape and the modal wavefront.
Application of this equation to the $r^4$-term in Eq. (\ref{eq:z}) for the LG$_{p\ell}$ modes yields
\begin{equation}
\label{eq:asphere}
    \Delta L_{\rm asp} 
    = - \frac{\langle \psi | r^4 | \psi \rangle}{8hR_m^2} 
    = - f(p,\ell) \frac{\gamma^4_z}{8hR_m^2} \,,
\end{equation}
where $\gamma_z^2 = \gamma_0^2 R_m/(R_m-L)$ is the mean square beam diameter at the concave mirror. The polynomial $f(p,\ell)$ introduced in refs. \cite{Yu1984,VanExter2022} is
\begin{eqnarray}
\label{eq:f(p,l)}
    f(p,\ell) & = & \frac{3}{2}(N+1)^2 - \frac{1}{2}(\ell^2-1) \,, \\
    \frac{3}{4} f(p,\ell) & = & f_{\rm asp}(N) - \frac{3}{8} \left( \ell^2 - \langle \ell^2 \rangle \right) \,,
\end{eqnarray}
with $f_{\rm asp}(N) = (N+1)^2+\frac{1}{2}$. 
The brackets in $\langle \ell^2 \rangle_\ell = N(N+2)/3$ indicate averaging over all $\ell$ values, i.e. over all $(N+1)$ modes in the $N=2p+\ell$ group.
This intriguing rewrite is motivated by our wish to separate the average effect $f_{\rm asp}(N)$ from the finestructure, and to later compare aspheric and nonparaxial effects; see Eq. (\ref{eq:nonparaxial}). We combine these results in a new equation
\begin{eqnarray}
\label{eq:fine-Gauss}
    \Delta L_{\rm asp} & = & \Ls \, [\frac{3}{8} \ell^2 - \frac{3}{8} \langle \ell^2 \rangle_\ell - f_{\rm asp}(N)] \,, \\
\label{eq:Ls}
    \Delta L_s & = & \frac{L}{6k^2h(R_m-L)} \,, \\
    G(L) & \equiv & \frac{\Ls}{\Ln} = \frac{L}{3h} \frac{R_m}{(R_m-L)} = \frac{\tilde {p} L}{3(R_m-L)} 
\end{eqnarray}
The parameter $\tilde{p}$ was used in ref. \cite{Koks2022b} to quantify the $r^4$- contribution to the mirror shape as $(1-\tilde{p})r^4/8R_m^3$ term, with $\tilde{p}=0$ for a sphere and $\tilde{p}=1$ for a paraboloid. 

Equation (\ref{eq:fine-Gauss}) neatly separates the average shift of the resonance length from the fine structure.
This average shift modifies the resonance condition to 
\begin{equation}
\label{eq:shifted}
    \langle L_j \rangle_\ell \approx \left[ q + (N+1) \frac{\chi_0}{\pi} \right] \frac{\lambda}{2} - f_{\rm asp}(N) \Ls(L) \,,
\end{equation}
and hence changes the transverse-mode spacing into 
\begin{eqnarray} 
\label{eq:DLtrans}
     \Delta L_{\rm trans}(q,N) & \equiv & L(q,N+1)-L(q,N) \nonumber \\
     & \approx & \Delta L_{\rm par} - (2N+3) \Ls(L) \,.
\end{eqnarray}
Spherical aberration with $\Delta L_s(L) > 0$ thus reduces the distance between consecutive transverse modes, or $N$-groups, and makes them non-equidistant.
A more practical way to describe this effect is through the introduction of the experimental Gouy phase $\chi_{{\rm exp}}=\chi_{{\rm exp}}(L,N) \equiv k \Delta L_{\rm trans}(q,N)$. 
After substitution of Eq. (\ref{eq:DLtrans}) and a Taylor expansion based on $\Ls(L) \ll \Delta L_{\rm par}$ we obtain 
\begin{equation} 
\label{eq:Gouy_scaling}
    \sin^2{\chi_{{\rm exp}}} \approx \frac{L}{R_m} \left[ 1 - \frac{1}{kh} \sqrt{\frac{L}{R_m-L}} (1+\frac{2}{3}N) \right] \,,
\end{equation} 
We thus find that $\sin^2[\chi_{{\rm exp}}(L,N)]$ is not strictly proportional to $L$, as in Eq. (\ref{eq:Gouy2}), but sublinear for a cavity with a Gaussian mirror.
The relative difference depends on the mirror depth $h$ via the factor $1/kh$ and is more prominent for high-order modes.
Equations (\ref{eq:DLtrans}) and (\ref{eq:Gouy_scaling}) also predict that consecutive $N$-groups are not precisely equidistant and that their spacing decreases with $N$.

Equations (\ref{eq:DLtrans}) and (\ref{eq:Gouy_scaling}) predict that the transverse-mode spacing in a cavity with a Gaussian mirror is less than in a cavity with a spherical mirror with the same $R_m$.
The reason is that Gaussian mirrors are less curved and provide less confinement, which tends to increase the size of the resonant modes. 
Appendix B attempts to quantify this effect by using $w_0$ as an adjustable parameter to minimize the mismatch between the modal wavefront and the mirror shape.
For the $N=0$ mode, it predicts that the effective radius of curvature of the matched mode at the curved mirror is 
\begin{equation} 
\label{eq:Reff}
    \frac{1}{R_{\rm eff}} 
    \approx \frac{1}{R_m} \left( 1 - \frac{1}{kh} \sqrt{\frac{L}{R_m-L}} \right) \,.
\end{equation}
This is identical to the $N=0$ result of Eq. (\ref{eq:Gouy_scaling}). 
For higher-order modes, this minimization yields results that are similar to Eq. (\ref{eq:Gouy_scaling}), but not perfectly identical as they also have a mild $\ell$-dependence.

For completeness, we note that one can also modify the effective mode size and curvature through small admixtures of $\Delta p = \pm 1$ modes with the same fixed $w_0$ and the same $\ell$ (imposed by rotation symmetry), i.e. through mode coupling with modes with different $(q,N)$ values.
This approach yields identical results.
In principle, small admixtures of $\Delta p = \pm 2$ modes can even modify the shape of the modal wavefront by an $r^4$-term, instead of an $r^2$ term.
However, the latter admixtures are only effective if these modes are sufficiently loss-free and not accidentally frequency degenerate with the original mode \cite{Benedikter2015,Benedikter2019,Koks2022a}.

Appendix \ref{sec:appendix-Reff} also describes how the remaining mismatch $\Delta z(x,y) = z_{\rm wave} - z_{\rm mirror}$ between wavefront and mirror shape introduces an intensity loss per roundtrip equal to \cite{Bennett1992} 
\begin{eqnarray}
\label{eq:scatter}
    {\rm A}_{\rm scatter} & = & \int [2k\Delta z(x,y)]^2 I(x,y) dxdy\,, \\
\label{eq:loss1}
    & \approx & \left( \frac{L}{4kh(R_m-L)} \right)^2 \,. 
\end{eqnarray}
An alternative equation, also derived in appendix \ref{sec:appendix-Reff}, attributes the intensity loss to some form of clipping loss.
Using the deflection point, $d^2 z_m/dr^2 =0$, of the Gaussian shape, somewhat ad hoc, as the clipping radius, one obtains 
\begin{equation}
\label{eq:loss2}
    A_{\rm deflection} \approx \exp{\left( -hk\sqrt{\frac{R_m-L}{L}} \right)} \,.
\end{equation}
A similar equation was already derived by Hunger et al. \cite{Hunger2010}. 
As ad hoc criterion for clipping, they use a radius were the mirror profile $z_m(r)$ is at $1/e$ of its maximum depth, instead of the $1/\sqrt{e}$ criterion that we use. 
This alternative criterion introduces a factor $2$ in the exponent of Eq. (\ref{eq:loss2}). 

Equations \eqref{eq:loss1} and \eqref{eq:loss2} both link loss to the limited mirror depth $h$ relative to $\lambda$, but their $L$-dependence and predicted values are different. 
For a typical case $L=R_m/2$ and $hk=\pi$, Eq. (\ref{eq:loss1}) predicts $A \approx 6 \times 10^{-3}$ while Eq. (\ref{eq:loss2}) predicts $A \approx 4 \times 10^{-2}$.
These losses affect the cavity finesse $F$ via
\begin{equation}
\label{eq:Finesse_from_loss}
    \frac{1}{F} = \frac{1}{F_0} + \frac{A}{2\pi} \,,
\end{equation}
where $1/F_0 = (1-{\cal R})/\pi$ for a cavity with two identical ideal mirrors with intensity reflection ${\cal R}$. 
The comparison between theory and experiment can hopefully decide which of the two equations is most useful. 
Both descriptions are highly speculative though, being based on rough assumptions only. 

If one is not interested in approximate analytic equations and instead wants exact results that include the detailed structure of the mirror, the approach described by Benedikter et al. \cite{Benedikter2015} should be taken, even though that approach does not include nonparaxial or vector effects. 
They introduce an extended set of modes, up to and beyond order $N=20$, and model the roundtrip evolution in this basis as a $400 \times 400$ matrix that they solve analytically. 
The obtained results are impressive and able to predict both the observed dips in the finesse close to frequency-degenerate points and the general drop in finesse at cavity length $L \geq R_m/2$.


\subsection{Nonparaxial effects}
\label{sec:nonparaxial}

The above calculation used a paraxial description of optical propagation. 
However, nonparaxial effects also modify the cavity resonances by themselves. 
Furthermore, these effects are intrinsic and unavoidable. 
For cavities with ideal spherical mirrors,  
the nonparaxial effects modify the resonant lengths by \cite{VanExter2022,Koks2022b}
\begin{eqnarray}
\label{eq:nonparaxial}
    \Delta L_{\rm non} & = & \Ln \left[-(\ell\cdot s+1)-\frac{3}{8}\ell^2+f_{\rm non}(N)\right] \nonumber \,, \\
\label{eq:Ln}
    & = & \Ln \left[-\ell\cdot s - \frac{3}{8} \ell^2 + \frac{3}{8} \langle \ell^2 \rangle_\ell - \frac{1}{2} \right] \,, \\
    \Ln & = & \frac{1}{2 k^2 R_m} \,, 
\end{eqnarray}
in their preferred eigen basis of vector LG modes. 
For cavities with other mirror shaped, the competition between mirror-shape and nonparaxial effects is not straightforward and requires a matrix description when these effects prefer different eigen bases; see Sec. \ref{sec:coupling}. 

The first term in Eq. (\ref{eq:nonparaxial}) contains the product of the orbital angular momentum (OAM) $\ell \geq 0$ and the photon spin $s = \pm 1$.
It describes optical spin-orbit coupling \cite{Bliokh2015} and originates from the boundary condition of the optical vector field at the curved mirror \cite{VanExter2022,Davis1984}.
The $\ell^2$ term originates from a $k_\perp^4$ contribution to the nonparaxial propagation and a $r^4$ contribution to the optical phase front \cite{VanExter2022,Yu1984}. 
The $f_{\rm non}(N) = (N^2+2N+4)/8$ term, which describes a general shift of all modes within each $N$ group, has the same origin but also includes a nonparaxial correction to the Gouy phase. 
The final rewrite 
shows that this general shift is small and independent of $N$. 
As a subtle modification, the radius $R_m$ in Eq. (\ref{eq:Ln}) should probably be replaced by the effective radius $R_{\rm eff}$ when the mode feels a large part of the mirror, because the latter determines the beam size and opening angle, and hence the nonparaxiality.

\subsection{Overview and scaling of all corrections}
\label{sec:overview}

\begin{table*}[htb]
    \centering
    \begin{tabular}{|c|c|c|c|c|c|}
    \hline 
     Contribution & Symbol & Operator form & Preferred basis & Magnitude $\Delta \tilde{L}_j = 2 \Delta L_j/\lambda$ & Approx. scaling\\ \hline \hline
     Paraxial & $(N+1)\Delta L_{\rm par}$ & $r^2$ and $k_\perp^2$ & degenerate & $(N+1)\chi_0(L)/\pi$ & $\propto \sqrt{L}$  \\ \hline \hline
     Anistropic (mirror) & $\Delta L_{\rm ani}$ & $x^2-y^2$ & scalar HG$_{mn}$ & $(m-n) \ea \tan{\chi_0(L)}/2\pi$ & $\propto \sqrt{L}$ \\
     Aspheric (mirror) & $\Delta L_{\rm asp}$ & $r^4$ & scalar LG$_{p\ell}$ & $\frac{3}{8} G(L) \ell^2 /2\pi kR_m$ & $\propto{L}$ \\ 
     Residual (mirror) & $\Delta L_{\rm \rm rest}$ & - & - & - & -  \\ \hline 
     Nonparaxial (scalar) & $ \ell^2 \, \Delta L_{\rm non}$ & $r^4$, $r^2 k_\perp^2$, $k_\perp^4$ & scalar LG$_{p\ell}$ & $-\frac{3}{8} \ell^2 /2 \pi kR_m$ & constant \\ 
     Nonparaxial (vector) & $\ell \cdot s \, \Delta L_{\rm non}$ & $\vec{r} \otimes \vec{k}_\perp$ & vector LG$_{p\ell}$ & $- (\ell \cdot s+1)/2\pi kR_{\rm m}$  & constant \\ \hline 
    \end{tabular}
    \caption{Overview of contributions to the transverse-mode spectrum, with their names, symbols, operator forms, and additional properties. 
    The paraxial contribution has no preferred basis, the anisotropic contribution prefers the Hermite-Gaussian (HG$_{mn}$) basis, while the aspheric and nonparaxial contributions prefer the Laguerre-Gaussian (LG$_{p\ell}$) basis. 
    The residual mirror effects are combined in $\Delta L_{\rm \rm rest}$, which is supposed to be small. 
    The fundamental Gouy phase $\chi_0(L) = \arcsin{\sqrt{L/R_{\rm m}}}$ and $\tan{\chi_0(L)} = \sqrt{L/(R_{\rm m}-L)}$.}
    \label{tab:table}
\end{table*}    

Table \ref{tab:table} gives an overview of the various contributions to the resonant length with their main properties.
The magnitude of these effects are expressed as $\Delta \tilde{L} = 2\Delta{L}/\lambda$, where $\Delta \tilde{L} \approx 1$ corresponds to one free spectral range $\Delta L_{\rm long}$. 
The first row describes the paraxial contribution $\Delta L_{\rm par}$.  
The next block of rows describe three effects of the mirror shape on the resonance condition.
The anisotropic $\Delta L_{\rm ani}$ originates from the $x^2-y^2$ astigmatism of the mirror shape.
The anisotropic contribution prefers HG modes and is approximately proportional to $\sqrt{L}$, similar to $\Delta L_{\rm par} \propto \chi_0$.
The aspheric $\Delta L_{\rm asp}$ originates from an $r^4$ contribution to the mirror shape.
The aspheric contribution prefers LG modes and scales approximately with $L$. 
The residual mirror-shape effects, like the $x^4-y^4$ high-order anisotropy, are finally combined in a rest correction $\Delta L_{\rm rest}$ and assumed to be small.
The final block of rows describe the corrections due to nonparaxial propagation and reflection \cite{VanExter2022}.
These effects are divided in a scalar correction, due to nonparaxial propagation and modified nonparaxial wavefronts, and a vector correction, which acts as a spin-orbit coupling (see below). 
All nonparaxial effects scale with $\Delta \tilde{L} \propto \lambda/R_m$ and are independent of cavity length.  


    \begin{figure}[ht]
    \centering
    \includegraphics[width=7.9cm]{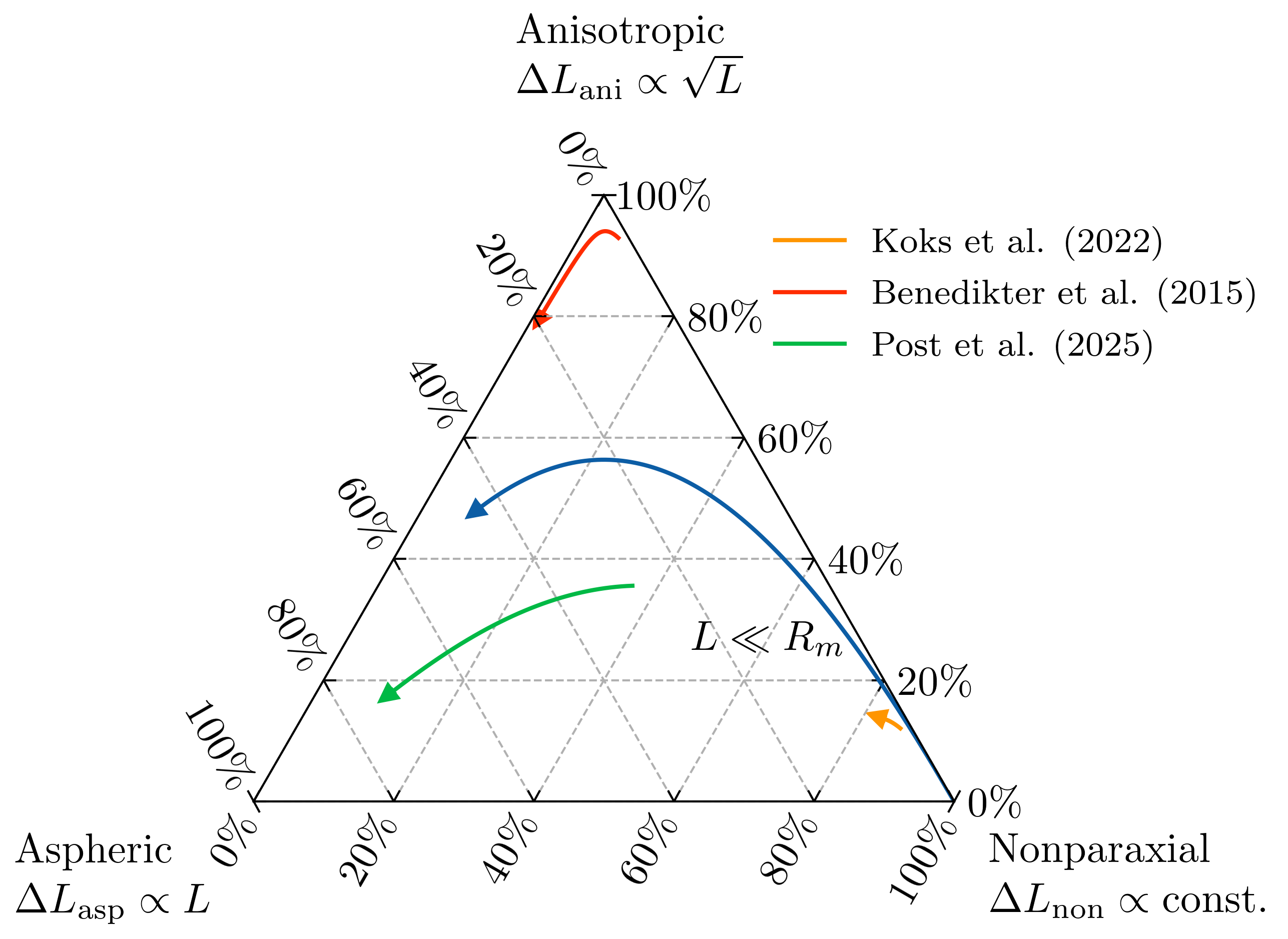}
    \caption{Visualization of the relative weight of three contributions to $\Delta L_{\rm fine}$ in a ternary plot, including their scaling with cavity length $L$. 
    The blue curve visualizes a typical shift of these relative weights from dominant nonparaxial at small $L$, via partially anisotropic to dominantly aspheric at large $L$. The other curves characterize the operational regime of the microcavities in \cite{Koks2022b} (orange), \cite{Benedikter2015} (red) and this work (green).}
    \label{fig:ternary}
    \end{figure}


Figure \ref{fig:ternary} summarizes the above discussion in a so-called ternary plot that show the relative weight of the various contributions to mode formation. 
Each cavity geometry can be represented by a single point inside this equilateral triangle, such that the distance to each leg measures the relative weight of each contribution (close to leg = far away from opposite corner = weak contribution of this effect).  
The bottom right corner represents all nonparaxial effects.
The opposite leg, between the ``aspheric" and ``anisotropic" corners, represent cavities with very weak nonparaxial effects. 
The model of Kleckner et al. \cite{Kleckner2010}, which uses scalar fields and neglects the scalar and vector nonparaxial corrections, is valid only in this regime.
The top corner of the triangle represents the anisotropic effect.
In cavities with dominant anisotropy, i.e. in the top corner of the triangle, the eigen modes are HG modes with modest polarization effects. 
In rotation-symmetric cavities, i.e. on the bottom leg of the triangle, the eigen modes are LG modes; either scalar LG (bottom left) when aspheric effects dominate or vector LG (bottom right) when nonparaxial effects dominate. 

The relative weight of the three 'corner' effects changes with cavity length $L$. 
The $\Delta L_j$ strength of the nonparaxial effect is constant, but the anistropic effect scales with $\sqrt{L}$ while the aspheric/Gaussian effect scales with $L$.
As a result, the representing point moves in the ternary plot when the cavity length is increased, for instance along the sketched blue curve. 


The dimensionless parameters $X$ and $G$ introduced above quantify the stengths of the anistropic and aspheric contribution relative to the nonparaxial effects. 
In dimensionless units, the longitudinal mode spacing $k \Delta L_{\rm long} \approx \pi$ and the transverse mode spacing $k\Delta L_{\rm trans} \approx \chi_0$. 
In the same units, the nonparaxial contribution to the fine structure $k \delta (\Delta L) = 2k \Delta L_n = 1/kR_m$ for the spin-orbit coupling between $\ell \cdot s = \pm 1$ modes, while the paraxial anistropic splitting $k \delta (\Delta L_{\rm trans}) = \eta_{\rm ani} \chi_0$, with $\chi_0 = \arcsin{\sqrt{L/R_m}}$.
The ratio between these two competing splittings is $\eta_{\rm ani} k \sqrt{LR_m} \approx X/4$ for $L \ll R_m$, which defines the dimensionless parameter $X$ in Eq. (\ref{eq:X}). 
The second nonparaxial effect, which scales with $(3/8)\ell^2$ instead of $\ell \cdot s$, can obviously also play a role. 
In cavities with Gaussian mirrors, this aspheric effect is enhanced by a dimensionless factor $G \approx (L/3h)R_m/(R_m-L)$ and can overtake the other effects for longer cavities with $G \gg 1$.

Let us estimate $X$ and $G$ for three different microcavities for an intermediate cavity length $L \approx R_m/4$, where $\chi_0 \approx \pi/6$.
The microcavity studied in ref. \cite{Koks2022b} was exceptional because it had very little anisotropoy $\eta_{\rm ani} = 0.6(2)$\% in combination with a small radius of curvature $R_m = 5.8(2)~\mu$m. 
At $\lambda = 0.633~\mu$m, this yields $2k \Delta L_n = 1/kR_m = 0.017$ for the mentioned spin-orbit coupling and $k \delta (\Delta L_{\rm trans}) = \eta_{\rm ani} \chi_0 = 0.003$, resulting in a modest anisotropic parameter $X/4\approx0.2$. 
The aspheric correction at $L \approx R_m/4$ was negligible at $G \approx -0.02$.
Hence, nonparaxial effects dominated, as confirmed by the observed transmission spectra and mode profiles. 
This cavity geometry therefore operated near the bottom right corner in the ternary plot (orange curve).

The microcavities studied in ref. \cite{Benedikter2015} had fiber mirrors with $R_x = 161~\mu$m and $R_y = 201~\mu$m, i.e. with a relatively large anisotropy $\eta_{\rm ani} = 11$\% and a relatively large average $R_m = 181~\mu$m. 
At $\lambda = 0.780~\mu$m, this yields $2k \Delta L_n = 1/kR_m = 0.0007$ and $k \delta (\Delta L_{\rm trans}) = \eta_{\rm ani} \chi_0 = 0.058$.
In these microcavities, the anisotropic effect strongly dominated with $X/4 \approx 80$. 
With a mirror depth as large as $h = 2.5~\mu$m, the aspheric effects were relatively modest at $G \approx 9$ for $L = R_m/4$.
As a result, the operating regime of this cavity is near the top corner of the ternary plot (red curve).

Finally, the microcavity studied in this paper has $\eta_{\rm ani} \approx 2.4$\% and $R_m \approx 13.6~\mu$m (see below). 
At $\lambda = 0.633~\mu$m, this yields $2k \Delta L_n = 1/kR_m = 0.0074$, $k \delta (\Delta L_{\rm trans}) = \eta_{\rm ani} \chi_0 = 0.0126$ and $X/4 = 1.7$. 
In these microcavities, nonparaxial effects play a significant role but do not dominate. 
Aspheric effects also play a role with $G \approx 2.5$ at $L \approx R_m/4$. 
In short, in our cavities all three mode-shaping effects compete in an intricate way, as illustrated by the path through the center of the ternary plot (green curve).

\subsection{Dynamic matrices}
\label{sec:coupling}

We already noted that the anisotropic correction prefers the HG basis whereas the anisotropic and nonparaxial corrections prefer the LG basis.
The eigen modes in cavities with both types of corrections will thus be superpositions of LG-modes or HG-modes.
These eigen modes and their eigen values follow from an eigen-mode analysis of the $(N+1) \times (N+1)$ dynamic matrices of each $(q,N)$ group.
Below, we will describe these matrices, which have also been called evolution or coupling matrices, for the $N=1, 2$ and 3 groups. 
This description extends the one presented in ref. \cite{VanExter2022}, because it includes the aspheric correction from the Gaussian mirror and presents the dynamic matrices both in the LG and HG basis.
It also separates the traceless part of the matrix from the general offset $f(N)$. 

The matrices presented below do not include the so-called hyperfine splitting, which can lift the pair-wise degeneracy of the modes.
Reference \cite{VanExter2022} attributes this hyperfine splitting to the phase difference in the reflection of $s$ and $p$-polarized light from a Bragg mirror (= DBR).
It then argues that this Bragg correction scales with the square of the angle of incidence on the mirror $\Theta_0^2 \propto 1/w_0^2 \propto 1/\sqrt{L(R_m-L)}$ and can even include a $\Theta_0^4$ contribution.
The hyperfine splitting should thus decrease with increasing cavity length. 
A second, much smaller, contribution to the hyperfine splitting in the form of anisotropic spin-orbit coupling will be discussed in appendix \ref{sec:appendix-Tomography}. 

The traceless part of the dynamic matrix for the $N=1$ group, in the $(1B,1A)$ basis of vector-LG modes, is \cite{VanExter2022}
\begin{equation}
\label{eq:N1}
    \Delta \hat{L}_{\rm LG} = \frac{\Delta L_n}{4} \begin{pmatrix} -4 & X \\ X & 4 \end{pmatrix} \Rightarrow \begin{pmatrix} - \Delta L_n & \Delta L_a \\ \Delta L_a & \Delta L_n \end{pmatrix} \,,
\end{equation}
where the hat indicates the matrix character of $\Delta \hat{L} = \Delta \hat{L}_{\rm fine}$. 
In the final matrix, we have substituted $\Delta L_a = \Delta L_n X/4$. 
The aspheric correction $\propto G$ is identical for both modes and hence only results in an overall shift of the $N=1$ modes.
The eigen values of the second matrix are the resonant lengths $L_j = \pm \sqrt{\Ln^2 + \La^2}$.
The eigen modes are linear superpositions like $\vec{\psi}_j = \cos{\beta} \vec{\psi}_{1B} + \sin{\beta} \vec{\psi}_{1A}$ with mixing angle $\beta = \arctan(\La/\Ln)$.  

A transformation of the dynamic matrix from the LG to the HG basis, using the transfer matrix $U  = [[1,-1],[1,1]]/\sqrt{2}$ and its inverse, yields
\begin{equation}
\label{eq:N1-HG}
    \Delta \hat{L}_{\rm HG} = \begin{pmatrix} \La & \Ln \\ \Ln & - \La \end{pmatrix} = 
    \frac{\Ln}{4} \begin{pmatrix} X & 4 \\ 4 & - X \end{pmatrix} .
\end{equation}
This matrix shows how nonparaxial corrections can deform the expect HG modes by admixing a HG mode with a different $(m, n)$ label.

The traceless part of the dynamic matrix for the $N=2$ group in the $(2B, 0, 2A)$ vector-LG basis is 
\begin{equation}
\label{eq:N2}
    \Delta \hat{L}_{\rm LG} = \frac{\Delta L_n}{4}  \begin{pmatrix} -10+2G & \sqrt{2} X & 0 \\ 
    \sqrt{2}X & 4-4G & \sqrt{2} X \\ 0 & \sqrt{2} X & 6+2G  \end{pmatrix} \,.
\end{equation}
The on-diagonal elements combine the spin-orbit splitting, $\mp 4(\ell \cdot s) = \mp 8$ for $\ell = 2$ and $s = \pm 1$, with the combined aspheric and nonparaxial contribution, $\frac{3}{2} (G-1) (\ell^2 - \langle \ell^2 \rangle_\ell) = (G-1)\{2,-4,2\}$, all in $\Delta L_n/4$ units.
The experiments presented in ref. \cite{Koks2022a} were performed in the regime $G \ll 1$ and modest $X$.
Their observations agreed with the prediction that the $2B$ modes occur at short cavity length (at -10), while the $0$ and $2A$ modes occur at longer cavity lengths and are closer together (at 4 and 6). 

The resonant modes will retain their vector LG character when $X$ is small but will mix for larger $X$.
This mixing will first occur for modes with comparable resonance lengths c.q. on-diagonal elements. 
For $G < 1$, these are the 0 and 2A modes, but for $G>1$ the outer $\ell = 2$ modes might be closer together. 
For very large $X$, it can be convenient to express the dynamic matrix in the HG-basis (HG$_{20}$, HG$_{11}$, HG$_{02}$). 
A transformation of the dynamic matrix from the LG to the HG basis, using the transfer matrix $U  = [[1,-\sqrt{2},-1],[\sqrt{2},0,\sqrt{2}],[1,\sqrt{2},-1]]/2$ and its inverse, yields 
$4 \Delta \hat{L}_{\rm HG}/\Ln = $ 
\begin{equation} \label{eq:N2_HG}
    \begin{pmatrix} 2X+1-G & 4 \sqrt{2} & 3-3G \\ 
    4\sqrt{2} & -2+2G & -4\sqrt{2} \\ 3-3G & -4\sqrt{2} & -2X+1-G  \end{pmatrix} \,.
\end{equation}
As on-diagonal elements we recognize the sequence $\{ 2X, 0, -2X \}$ expected when anisotropy dominates. 
But the other matrix elements, which originate from the aspheric and nonparaxial corrections, typically cannot be neglected and often even dominate.

The traceless part of the dynamic matrix for the $N=3$ group in the $(3B, 1B, 1A, 3A)$ vector-LG basis $4 \Delta \hat{L}_{\rm LG}/\Delta L_n =$
\begin{equation}
    \begin{pmatrix} -18+6G & \sqrt{3}X & 0 & 0 \\  \sqrt{3}X & 2-6G & 2X & 0 \\ 0 & 2X & 10-6G & \sqrt{3}X \\ 0 & 0 & \sqrt{3}X & 6+6G  \end{pmatrix} \,.
\end{equation}
As on-diagonal elements we recognize the sequence $\{-12, 4, 4, 12 \}$ from spin-orbit coupling in combination with $\pm 6(G-1)$ terms from the combined aspheric and various nonparaxial corrections. 
As off-diagonal elements we recognize the effects of anisotropy, already described in refs. \cite{VanExter2022,Koks2022b}. 

A conversion of the $N=3$ dynamic matrix from the vector-LG to the HG basis results in 
\begin{widetext}
\begin{equation}
\label{eq:HG3}
4 \Delta \hat{L}_{\rm HG}/\Ln =  
\begin{pmatrix} 3X + 3(1-G) &  4\sqrt{3} & 3\sqrt{3}(1-G) & 0 \\  4\sqrt{3} & X - 3(1-G) & -8 & 3\sqrt{3}(1-G)  \\ 3\sqrt{3}(1-G)  & -8 & -X - 3(1-G) & 4\sqrt{3} \\ 0 & 3\sqrt{3}(1-G) & 4\sqrt{3} & -3X + 3(1-G)  \end{pmatrix} \,.
\end{equation}
\end{widetext}
In the final matrix, we again recognize the on-diagonal contribution from the anisotropy, the next to diagonal elements from the spin-orbit coupling, and other elements that originate from nonparaxial and aspheric effects and hence scale with $(1-G)$.
Below we will show that the vector-HG modes in each basis set have alternating linear polarizations.  
As a result, spin-orbit coupling, which produces the next to diagonal element, will mix modes with orthogonal polarizations while the $(1-G)$ effects, which act on the scalar field, naturally only mix modes with the same polarization.  

A reader interested in dynamic matrices of $N \geq 4$ groups could either consult ref. \cite{Koks2022b} for a simple version of the $N=4$ matrix, or perform the operator algebra described in ref. \cite{VanExter2022}, or contact the authors.
It simply doesn't make sense to add more dynamic matrices to the current paper. 

\subsection{Symmetry and polarization of mode}
\label{subsec:symmetry}

The optical polarization doubles the number of modes per $N$ group from $(N+1)$ scalar modes to $2(N+1)$ vector modes.
That the dynamic matrices presented above are still $(N+1) \times (N+1)$ matrices is a result of mirror symmetry.
The cavity geometry that we considered has inversion symmetry along the optical $z$ axis and mirror symmetry in the $xz$- and $yz$-planes, where the $x$ and $y$ directions are set by the astigmatic mirror.
The cavity eigen modes must also obey these symmetries and can thus be divided in modes that are either even (+) or odd (-) with respect to the mirror action in the $xz$ axis. 
The symmetry under the other mirror action automatically follows, because the combination of both mirror actions is an inversion that introduces identical phase factor $(-1)^{N+1}$ to all modes in the $N$-group. 
Symmetry thus allows one to split the $2(N+1)$ vector modes in two sets of $(N+1)$ modes with either even (+) and odd (-) $xz$ mirror symmetry. 
And because modes with different symmetry do not couple, the evolution within each set is described by its own $(N+1) \times (N+1)$ dynamic matrix.   
Furthermore, the dynamic matrices for the two sets of modes are identical when hyperfine effects are negligible \cite{VanExter2022}.
Hence the eigen modes with + and - symmetry are pairwise degenerate. 

The assignment of $+/-$ labels is easy for the $x$ and $y$-polarized vector HG$_{mn}$ modes.
The $x$-polarized HG$_{mn}$ modes have $+$ symmetry for even $n$ and $-$ symmetry for odd $n$. 
The $y$-polarized HG$_{mn}$ modes have the opposite combinations.
As a consequence, the HG-vector modes in each symmetry set alternate between $x$ and $y$ polarization.
For example, the HG-vector modes with + symmetry in the $N=3$ group are, from top to bottom, the $x$-polarized HG$_{30}$, the $y$-polarized HG$_{21}$, the $x$-polarized HG$_{12}$, and the $y$ polarized modes. 
The assignment of $+/-$ labels for the vector-LG modes is more complicated and discussed in ref. \cite{VanExter2022}, where modes with identical $(p,\ell)$ labels are further divided in $\ell A+, \ell A-, \ell B+$ and $\ell B -$ modes depending on the alignment of spin and orbital momentum \cite{Yu1984}.

In general, $+$ modes can be converted into equivalent $-$ modes by a $90^\circ$ rotation of the local polarization. 
The $+$ and the equivalent $-$ mode can thus be written as 
\begin{eqnarray}
    \vec{E}_+(x,y) & = & E_{e}(x,y) \vec{e}_x + E_{o}(x,y) \vec{e}_y \nonumber \,, \\
    \vec{E}_-(x,y) & = & - E_{o}(x,y) \vec{e}_x + E_{e}(x,y) \vec{e}_y \,,
\end{eqnarray}
with $E_{e}(x,-y) = E_{e}(x,y)$, $E_{o}(x,-y) = - E_{o}(x,y)$, such that 
modes with $+/-$ symmetry have a pure $x/y$-polarized field on the $x$-axis. 
We can further decompose these fields in the vector-HG basis as
\begin{eqnarray}
\label{eq:even-odd}
    E_{e}(x,y) & = & \sum_{i {\rm = even}} \alpha_i HG_{N-i,i}(x,y) \nonumber \,, \\
    E_{o}(x,y) & = & \sum_{i {\rm = odd}} \alpha_i HG_{N-i,i}(x,y) \,,
\end{eqnarray}
with amplitudes $\alpha_i$ and scalar-HG modes $HG_{mn}(x,y) = HG_m(x) HG_n(y)$.
All amplitudes $\alpha_i$ must have the same phase at the flat mirror, to produce a flat phase front, and can thus be taken real-valued at this mirror.
And they remain approximately in phase during propagating to the curved mirror, because they experience approximately the same Gouy phase lag $(N+1)\chi_0$. 

Finally, we consider the case where the cavity is excited with an input beam with some general polarization and spatial misalignment. 
At any resonance length, this beam will typically excite a linear superposition of two degenerate eigen modes of the form
\begin{equation}
    \vec{E}(x,y) = \beta_+ \vec{E}_+(x,y) + \beta_- \vec{E}_-(x,y) \,.
\end{equation}
The excitation amplitude $\beta_+$ and $\beta_-$ will obviously depend on the input polarization and alignment but will also be different for the different resonances, because their modal overlap with the input is different.
Despite this complication, we can still split the excited resonant mode in its $+$ and $-$ components by combining $\vec{E}(x,y)$ with $\vec{E}(x,-y)$ as
\begin{eqnarray} \label{eq:split_symm_Efield}
    E_x(x,y)+E_x(x,-y) &  = & 2 g\beta_+ E_{e}(x,y) \nonumber \,, \\ 
    E_y(x,y)-E_y(x,-y) &  = & 2 \beta_+ E_{o}(x,y) \nonumber \,, \\
    E_x(x,y)-E_x(x,-y) &  = & - 2 \beta_- E_{o}(x,y) \nonumber \,, \\ 
    E_y(x,y)+E_y(x,-y) &  = & 2 \beta_- E_{e}(x,y) \,, 
\end{eqnarray}
where the $(x,y)$ axes still correspond with the axes of the elliptical mirror. 
The redundancy in the above equations can serve as a check of this symmetrization procedure.

\section{Experimental setup}
\label{sec:setup}

\begin{figure}[ht]
    \centering
    \includegraphics[width=7.9cm]{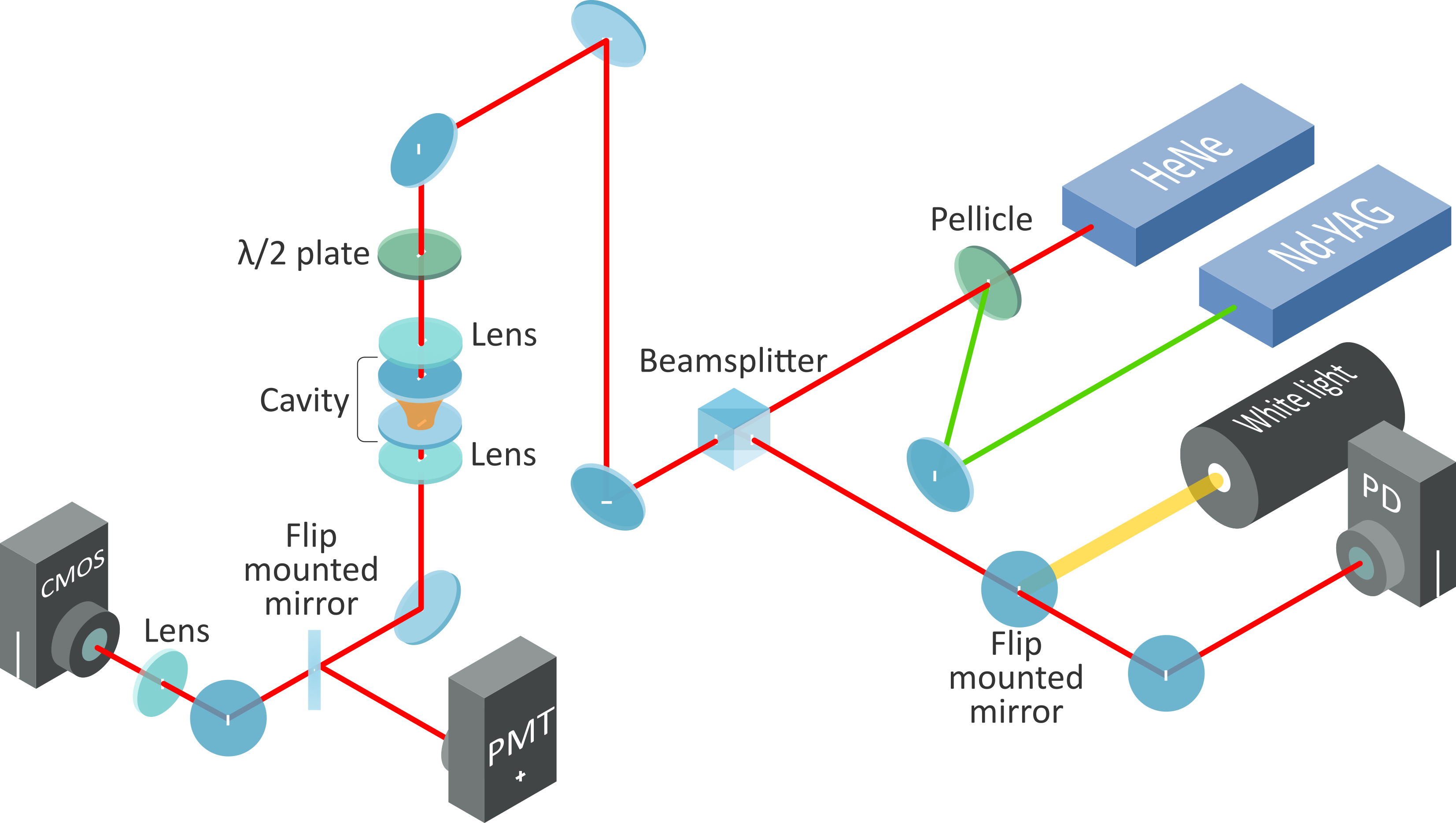}
    \caption{Schematic setup of the experiment used to characterize a microcavity (on the left) in various ways.  
    The HeNe-laser (633 nm) is used for measurements, while the Nd:YAG-laser (532 nm) and white light source are used for alignment and imaging. 
    Transmission and reflection spectra are measured with an photomultiplier tube (PMT) and a (normal) photodiode (PD), respectively. 
    Resonant mode profiles are observed with a polarization-resolving camera.}
    \label{fig:setup}
    \end{figure}

Figure \ref{fig:setup} shows the schematic setup of experiment. 
The microcavity shown on the left consists of a planar and a concave mirror. 
The planar mirror was produced by Laser Optik (Transmission $T = 1.4 \times 10^{-3}$ at $\lambda$ = 633 nm and $T = 1.0 \times 10^{-3}$ in the center of the stop at $\lambda_c$ = 610 nm). 
The concave mirrors were produced by Basel university and also coated by Laser Optik ($T \approx 1.0 \times 10^{-3}$ at $\lambda$ = 633 nm with $\lambda_c$ = 637 nm). 
These mirrors contained multiple concave structures, which were produced with the technique of CO$_2$ laser ablation on planar mesas \cite{Hunger2012,Najer2017,Najer2019}, 
with typical radii of curvature of 12 - 21 $\mu$m, mirror depth 0.15 - 0.3 $\mu$m, and anisotropy $\ea = 1.1 - 6.2 \%$ (see below).
The concave mirror is an H-DBR mirror that ends with the high-index medium, Ta$_2$O$_5$ ($n$ = 2.12), while the planar mirror is a L-DBR that ends with the low-index medium, SiO$_2$ ($n$ = 1.481), both on silica substrates. 
This distinction affects the positions of the nodes and anti-nodes of the intra-cavity field \cite{Koks2021}.

The upper concave mirror is fixed, while the bottom planar mirror is on a hexapod system that can be moved in six degrees of freedom with better than 0.1 $\mu$m and 0.1 mrad precision. 
We align the mirrors to the point where they are parallel and almost touch each other; the touching-point is referred to as ``touchdown". 
We scan the mirror position over $> 2~\mu$m distance with piezo actuators, relative to the position set with the hexapod. 

We inspect our microcavities with three different light sources (see Fig. \ref{fig:setup}).
We use a white light source for imaging and rough alignment and a Nd-YAG laser ($\lambda = 532$~nm) for further alignment, using the interference fringes visible between the flat parts of both mirrors.
We use a HeNe-laser ($\lambda = 633$~nm) for the final characterization, which involves (i) measuring the transmission and reflection spectra, defined as intensity versus cavity length, and (ii) observing the (angular) mode profiles under resonant transmission.  
All light sources are focused into the cavity via a microscope objective (40$\times$, $f = 5$~mm, N.A. = 0.60), while the transmitted light is collected with a f = 8 mm lens. 
An photomultiplier tube (PMT, Hamamatsu H5783) and a (normal silicon) photodiode (PD, Thorlabs PDA8A2) are used to measure the transmitted and reflected power respectively, while quickly scanning the cavity length with piezo voltage. 
A polarization-sensitive CMOS camera (FLIR Blackfly BFS-U3-51S5P-C) yields the intensity profile I ($\theta{_x}$, $\theta{_y}$) at the output lens and its polarization profile $\vec{e}$ ($\theta{_x}$, $\theta{_y}$) at different cavity lengths $L$, under slow scanning condition. 
This camera consists of a $1024 \times 1224$ array of super pixels, each comprising $2 \times 2$ arrays of pixels covered with horizontal, vertical, diagonal and anti-diagonal polarizers, which together yield the polarization profile but can't distinguish right-handed from left-handed polarization. 
The results from these two different analysis methods will be presented in two different sections, called `resonance spectra' and `resonant mode profiles'.



\section{Resonance Spectra}
\label{sec:spectra}

\subsection{Transmission Spectrum}
\begin{figure}
    \centering
    \includegraphics[width=7.9cm]{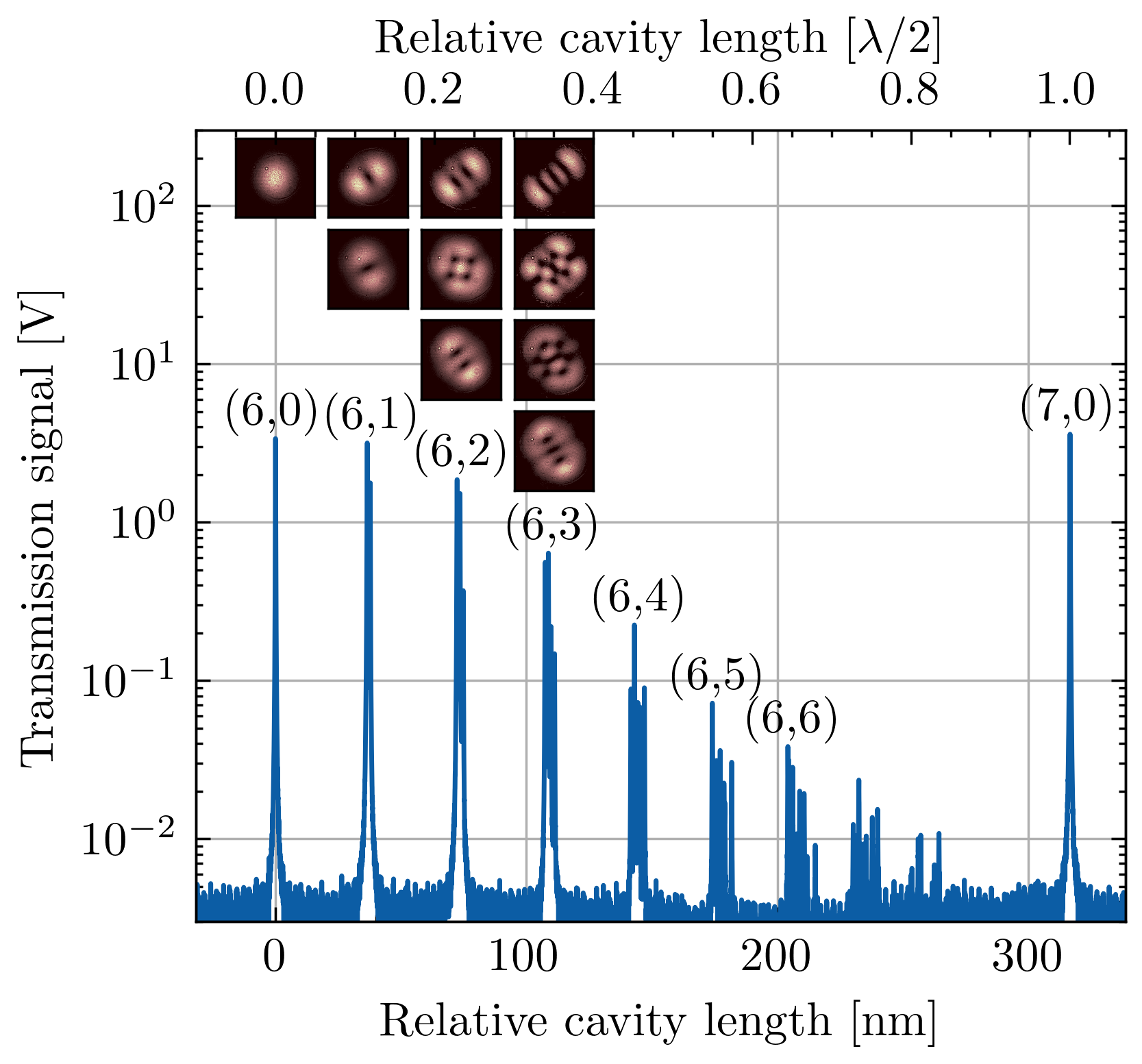}
    \caption{Transmission spectrum, expressed as PMT voltage versus relative cavity length in nm (bottom axis) and in units of $\lambda/2$ (top axis). Each group of transmission peaks is labeled by two quantum numbers $(q, N)$ and has a detailed spectral fine structure, visible in the spectrum and mode profiles (see insets).
    }
    \label{fig:transmission spectrum}
\end{figure}
Figure \ref{fig:transmission spectrum} shows the transmission spectrum of our cavity around small cavity length ($q=6$) for an on-purpose misaligned input. 
Each group of transmission peaks is labeled by its longitudinal mode number $q$ and its transverse order $N$ as $(q, N)$. 
The cavity length is measured relative to the fundamental ($N=0$) mode, using the free spectral range ($\approx 1.01\ \lambda/2$; see appendix \ref{sec:appendix-Calibration}) for scaling.
The values of the longitudinal mode number are deduced from the analysis presented in Sec. \ref{subsec:gouy_phase}. 
The semi-logarithmic vertical scale highlights the large dynamic measurement range and the existence of many transverse-mode groups $N$. 
The absolute peak transmission is about 10\% for the most prominent peaks but it can be much higher for matched incoupling (see below).
The observation that we only excite transverse modes over a finite $N$ range, despite the (significant) intentional misalignment, gives a first indication of the finite mirror depth $h$.  
Each $(q,N)$ group contains multiple peaks with slightly different resonance lengths. 
Below, it is shown how this so-called fine structure can be used to characterize the microcavity and the shape of its concave mirror.
Analysis of the mode profiles will be discussed in the next section. 
A first observation of these modes, as presented in the insets in Fig. \ref{fig:transmission spectrum}, already shows that they are not simple HG or LG modes and that their intensity patterns somewhat resemble the patterns predicted for the scalar Ince-Gaussian modes \cite{Bandres2004}. 
This suggests the presence of various effects competing in strength.

\begin{figure}
    \centering
    \includegraphics[width=7.9cm]{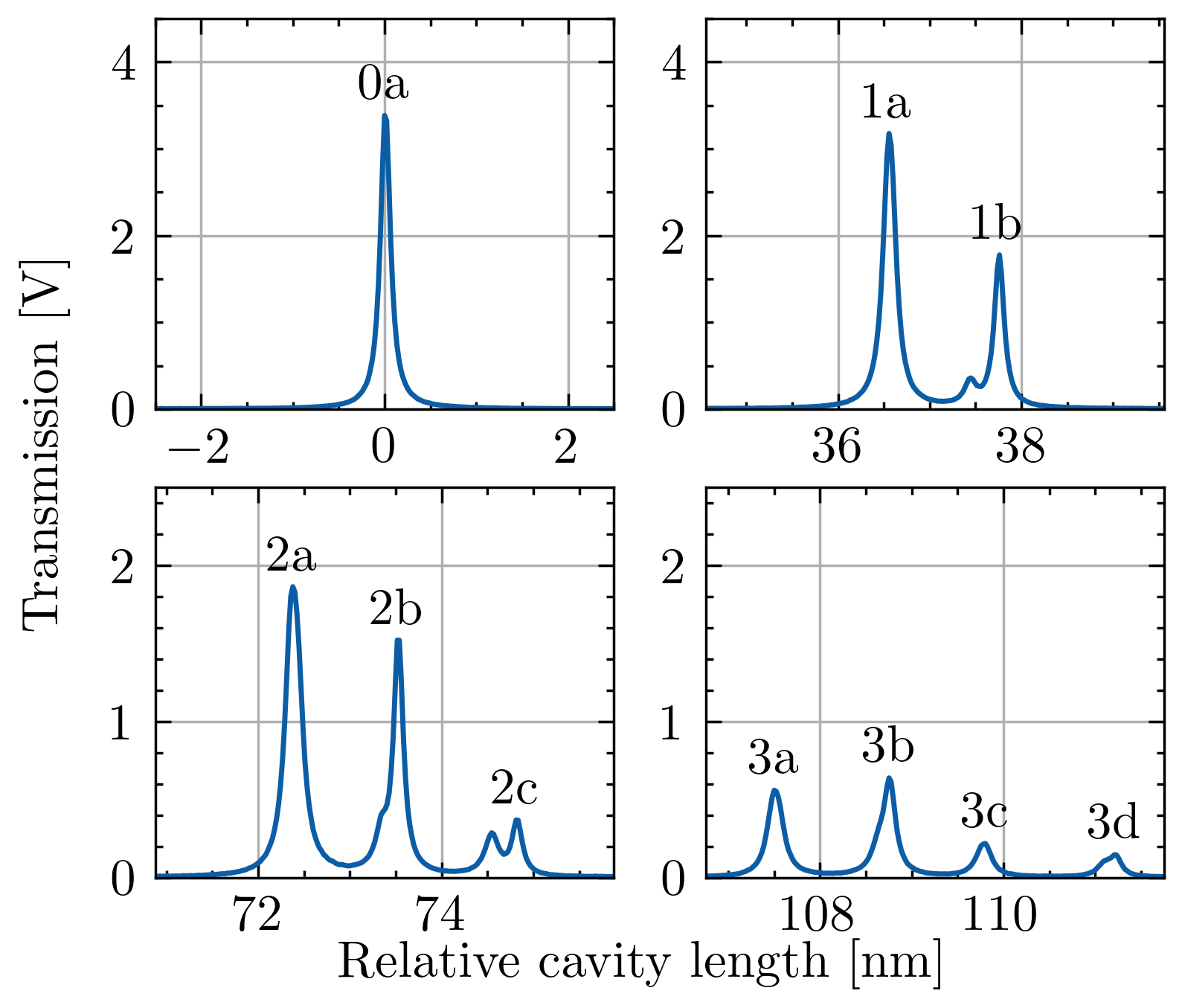}
    \caption{Detailed views of the fine structure in the $(q=6 ,N)$ groups in Fig. \ref{fig:transmission spectrum} for $N=0$ to 3. 
    Each group has $(N+1)$ resonant peaks, but some peaks ($1b, 2b, 2c)$ are split in two due to an observable hyperfine splitting.}
    \label{fig:FineStructure}
\end{figure}

Figure \ref{fig:FineStructure} shows enlarged views of the transverse mode groups up to $N=3$. 
These figures show that the resonance spectrum of each $(q,N)$ group is typically split into $N+1$ modes, the so called \textit{fine structure}. 
To distinguish these, the transverse-mode label will be appended alphabetically starting at `a' for the peak at shortest cavity length. 
In some cases, one can even resolve another splitting, which is referred to as the \textit{hyperfine} splitting \cite{VanExter2022, Koks2022b}. 
This is particularly evident in the 1b and 2c resonances, but can also be seen in the 2b resonance. 


Different alignments cause different magnitudes in the various peaks. Specifically, in the measurement of Figures \ref{fig:transmission spectrum} and \ref{fig:FineStructure}, the alignment of the laser deviates on purpose from perfect alignment with the fundamental mode in order to efficiently excite high-order modes. Measurements at different alignments (not shown in figure) have shown differences in magnitudes of the various transverse modes, but the positions and widths of the peaks did not depend on the alignment.

The remainder of this section presents itself as a five-step recipe for experimentalists to characterize their cavity based on its transmission spectrum. 
The section ends with measurements and analyses of the finesse, the optical polarization, and reflection spectra.


\subsection{Step 1: Gouy phase}
\label{subsec:gouy_phase}
A measurement of the Gouy phase $\chi_0(L)$ and its variation with $L$ allows one to determine the radius of curvature, according to Eq. \eqref{eq:Gouy2} for a spherical mirror or Eq. \eqref{eq:Gouy_scaling} for a Gaussian mirror. 
It further allows one to match correct values of $q$ to the measured resonances in the iterative manner discussed in Appendix \ref{sec:appendix-Calibration}.
Figure \ref{fig:Gouy_phase} shows $\sin^2{\chi_0}$ deduced from the ratio of the measured transverse- and longitudinal-mode splittings for the $N=1$ versus the $N=0$ group according to Eq. \eqref{eq:definition-Gouy}. 
The black solid line presents a linear fit to the data at low cavity lengths ($q \leq 11$), using a guessed $q_{\rm min} = 5$. 
The linear fit curve crosses the horizontal axis at $\Delta L=-1.36\ \lambda/2$. 
Appendix \ref{sec:appendix-Calibration} argues that the expected axis crossing is $L = - \Delta \tilde{q} -\Delta L_{\text{\rm pen}}$, with integer $\Delta \tilde{q}$ and expected penetration depth 
$\Delta L_{\text{\rm pen}} \approx 0.38\ \lambda/2$. 
This offset therefore suggests the lowest measured longitudinal mode is actualy $q_{\text{min}}=6$ with a penetration depth of $\Delta L_{\text{\rm pen}} = 0.36\ \lambda/2$.
The inverse of the slope of the fit yields a radius of curvature of $R = 15.8(5)~\mu$m. 


\begin{figure}
    \centering
    \includegraphics[width=7.9cm]{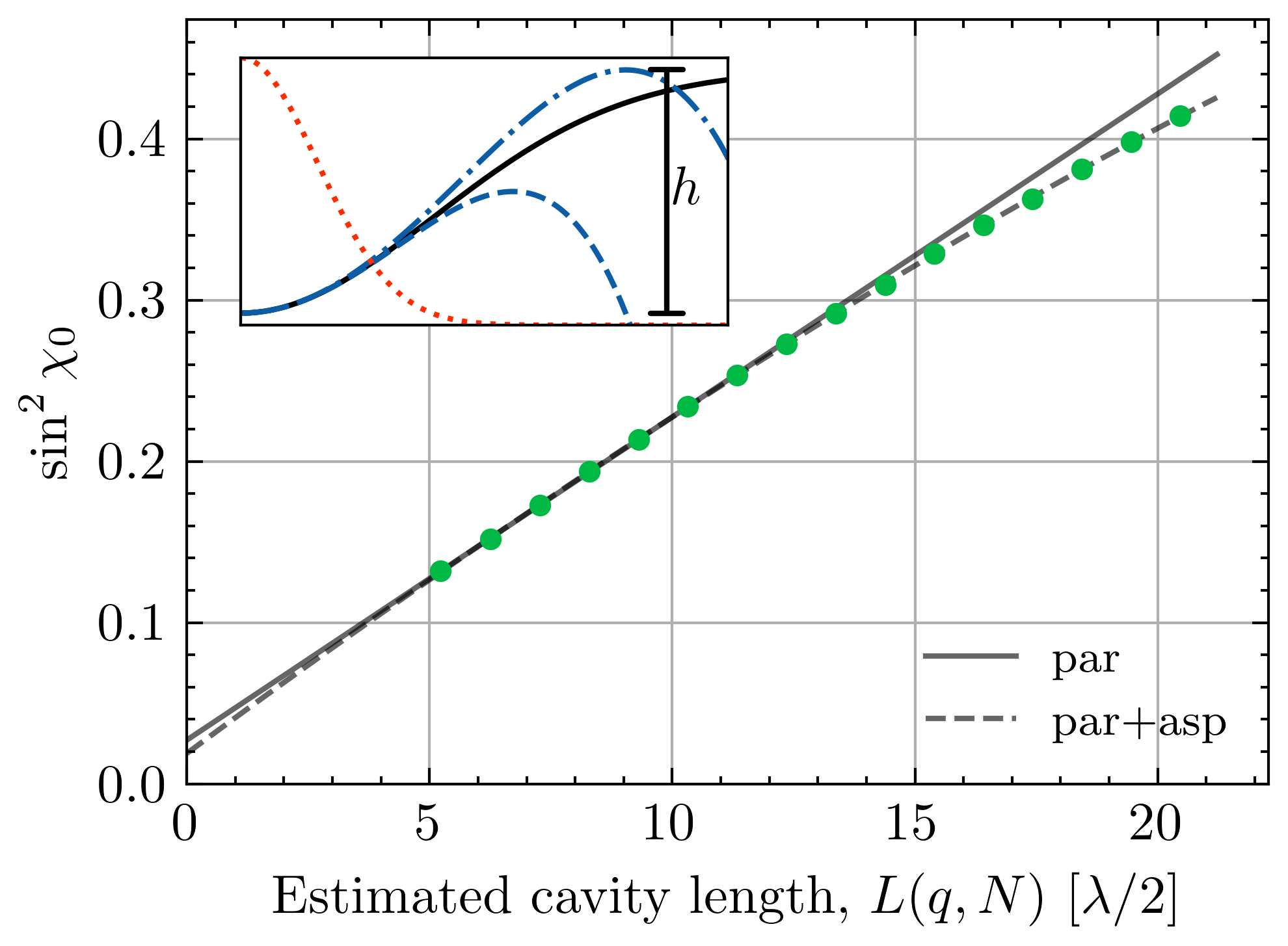}
    \caption{sin$^2(\chi_0)$, deduced from measured transverse-mode spacing $\propto \chi_0(L)$, versus cavity length $L$. 
    The black solid line is a linear fit at short cavity length $L$. 
    The black dashed curve is a fit based on Eq. (\ref{eq:Gouy_scaling}). 
    The inset shows $z_m(r)$ of a perfect Gaussian-shaped mirror with depth $h$ (black solid curve) and two Gaussian expansions up to $r^4$ according to Eq. \eqref{eq:z} (blue dashed curves), for $h_4 = h$ (lower curve) and $h_4=2h$ (upper curve), with the same $R_m$. The red dotted curve shows the intensity profile $I(r) \propto e^{-r^2/\gamma^2}$ of the fundamental mode at $L=R_{m}/2$.}
    \label{fig:Gouy_phase}
\end{figure}

At longer cavity lengths, the data deviates from the linear curve, suggesting the presence of an aspheric contribution according to Eq. \eqref{eq:Gouy_scaling}. 
The black dashed curve shows how Eq. \eqref{eq:Gouy_scaling} neatly fits the data for $R_m = 13.6(5)\ \mu$m and $h_4=0.61(6)\ \mu$m, keeping in mind that this $h_4$ is merely a convenient fit parameter for the amount of spherical aberration. The inset in Fig. \ref{fig:Gouy_phase} visualizes how $h_4$ does not necessarily equal the mirror depth $h$, but may nevertheless provide an order-of-magnitude estimate of it.
The black dashed curve crosses the horizontal axis at $\Delta L=-0.83\ \lambda/2$.
This suggests that the lowest measured longitudinal mode might actually be $q_{\text{min}}=5$, instead of 6, with a penetration depth of $\Delta L_{\text{\rm pen}} = 0.83(5)\ \lambda/2$. 
The horizontal scale in Fig. \ref{fig:Gouy_phase} is such that the lowest measured mode is indeed at $L\approx5\ \lambda/2$.


\begin{figure}
    \centering
    \includegraphics[width=7.9cm]{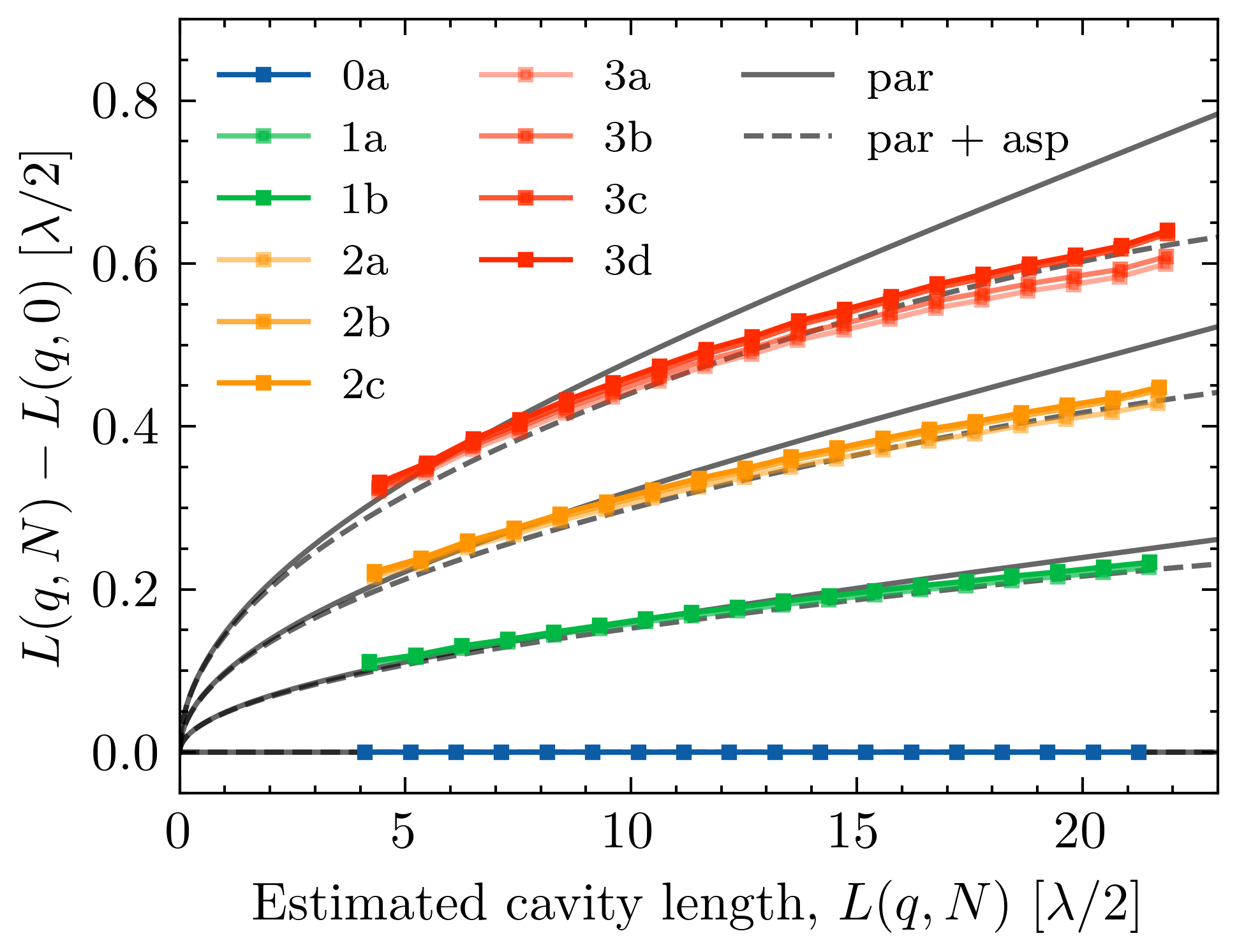}
    \caption{Measured mode spacings (colored curves) between various $(q, N)$ modes and the $(q,0)$ mode. The black solid curves show the mode spacing predicted by paraxial theory, while the black dashed curves present predictions that include the aspheric contribution for $R_m = 13.6~\mu\text{m}$ and $h_4 = 0.61~\mu\text{m}$. The measurements show the fine structure splittings over $(N+1)$ modes as well as the overall deviation from paraxial theory for increasing $N$.
    }
    \label{fig:full_theory_peaks}
\end{figure}

Figure \ref{fig:full_theory_peaks} presents the mode spacing $L(q,N)-L(q,0)$ versus cavity length. 
The $N=1$ data are identical to the ones presented in Fig. \ref{fig:Gouy_phase} and a fit of these data with Eq. \eqref{eq:DLtrans} yields identical fit parameters. 
The black solid curves show the paraxial predictions $N\chi_0(L)/\pi$ in $\lambda/2$ units. 
The black dashed curves show predictions based on Eq. \eqref{eq:DLtrans}.
The excellent agreement between observations and predictions for the $N=2$ and $N=3$ modes effectively serves as a check of the $(2N+3)$ scaling of the aspheric effect predicted by Eq. \eqref{eq:DLtrans}.
The closely spaced curves in Fig. \ref{fig:full_theory_peaks} also show the presence of fine structure in all $(q, N)$ spectra, as discussed below.

We conclude that the scaling of the Gouy phase with cavity length is well understood, including its dependence on the $r^4$ shape of the mirror via $h_4$, used as fitting parameter in the Taylor expansion in Eq. \eqref{eq:Gouy_scaling}. 
The radius of curvature found from the linear fit ($R=15.8~\mu$m) can be understood as the effective radius $R_{\rm eff}$ according to Eq. \eqref{eq:Reff}.
For further analyses we will continue with $q_{\text{min}}=5$ and $R_m=13.6(5)~\mu$m as convincingly provided by the aspheric corrected fit.


\subsection{Step 2: Mirror depth from planar modes} \label{subsec:MirrorDepth}
\begin{figure}
    \centering
    \includegraphics[width=7.9cm]{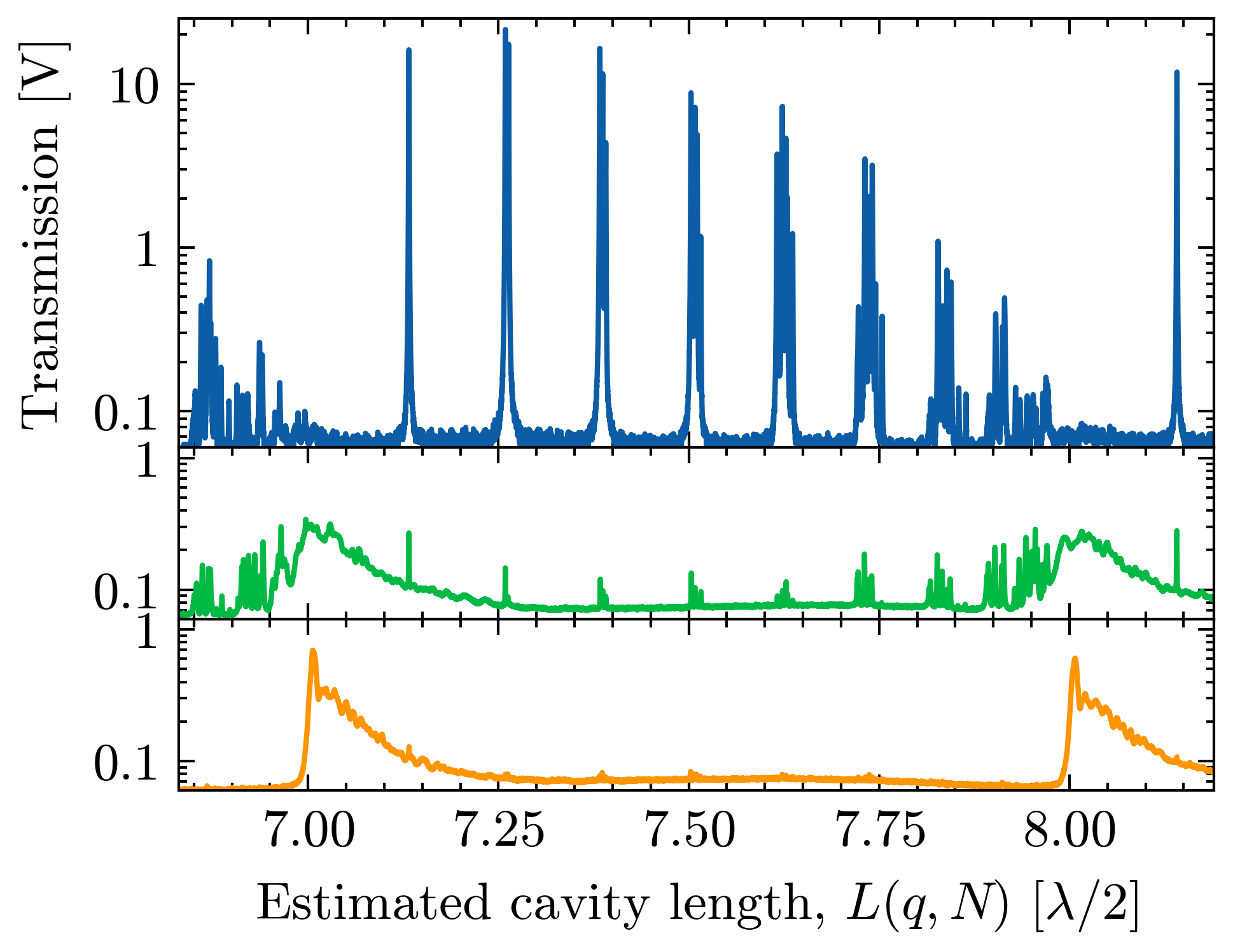}
    \caption{
    Transmission spectra for increasing misalignment. While the blue curve in the top row shows a slight misaligned plano-concave cavity spectrum, the more misaligned spectra in the middle and bottom row shows the rise of a planar spectrum. The comparison between these modes can be  used to determine the mirror depth $h$.
    }
    \label{fig:cav_to_panar}
\end{figure}

The mirror depth $h$ can be roughly estimated from the extent to which higher-order ($q,N$) modes can be excited upon misalignment. 
The data in Fig. \ref{fig:transmission spectrum} already suggest that the mirror depth $h$ is hardly larger than the free spectral range $\approx \lambda/2$.
Figure \ref{fig:cav_to_panar} shows a zoom-in of the transmission spectrum measured for increasing misalignment.
The broad features in the green curve originate from the excitation of planar modes, modes that remain visible for beam displacement beyond the mirror size. In the orange curve the planar mode is dominantly excited, while the cavity modes have almost vanished. The asymmetric shape of the curve can be understood from the planar resonance condition $L \cos{\theta} = q' \lambda/2$.
We compare these planar modes with the equivalent planar cavity mode, obtained by extrapolating the observed $(q,N)$ cavity modes to a fictional $N=-1$ mode located exactly at $L(q,-1)= q\ \lambda/2$, and can interpret their difference as the mirror depth. 
This comparison, which has been repeated for misalignment along several directions, is quite accurate and yields $h \approx 0.96(4)\ \lambda/2 \approx 0.30(1)\ \mu m$.

\subsection{Step 3: Fine Structure (qualitative)} \label{subsec:FineStructure}

\begin{figure*}
    \centering
    \includegraphics[width=\linewidth]{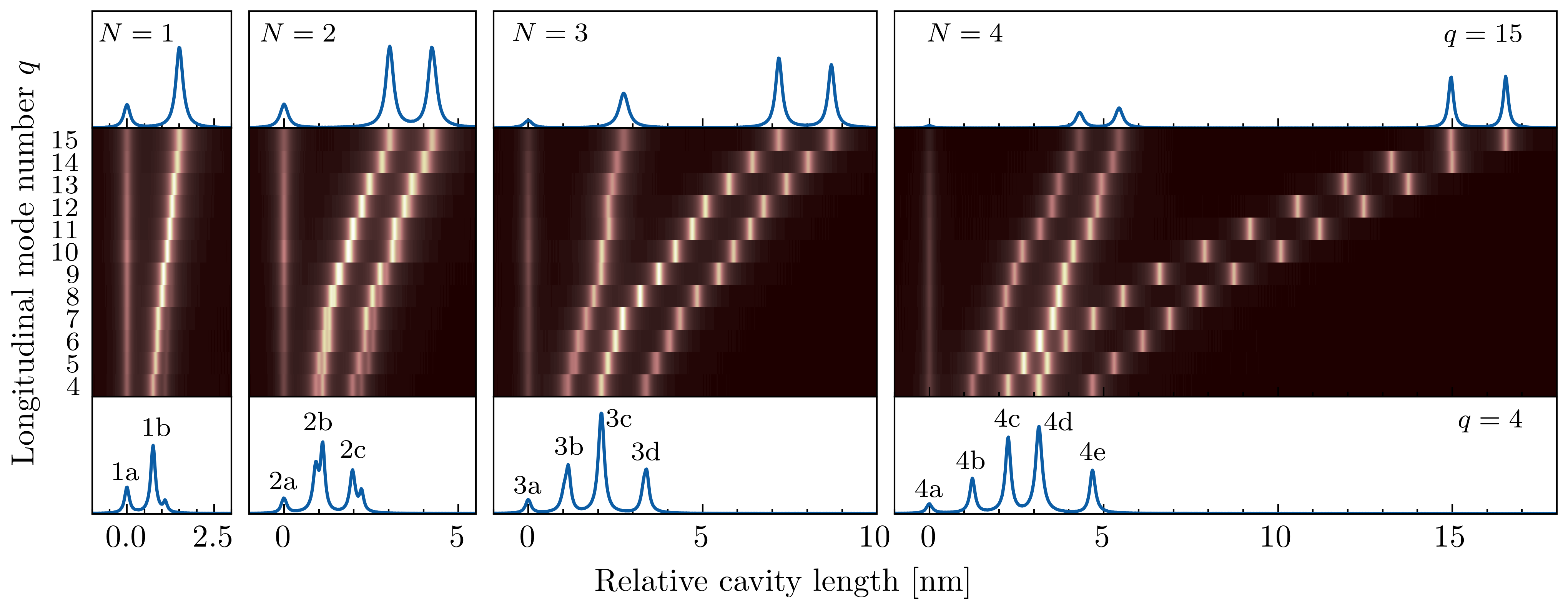}
    \caption{Fine structure for $N=1-4$ groups over a range of $q$ values. The peak at the shortest cavity length is now used as reference for the length scale of each $N$ group. The fine structure resonances are spaced equidistant at short cavity length ($q=4$). However, at larger cavity lengths the resonances organize themselves in pairs.}
    \label{fig:fine_structure_peaks}
\end{figure*}

Next, we consider the fine structure in the transmission spectra, qualitatively in step 3 and quantitatively in steps 4 and 5. 
Figure \ref{fig:fine_structure_peaks} shows the fine structure for different $N$ groups ($N=1,2,3,4$) over a range of $q$ values. 
Each spectrum now has been shifted such that the peaks at the shortest cavity length provide the reference length relative to which the positions of the other modes can be measured.

A first qualitative observation is that each $N$ group typically consists of $N+1$ peaks, as predicted by the nonparaxial theory.
Theory also predicts that all peaks should be doublets \cite{VanExter2022,Koks2022a}. 
Some of these doublets are actually split in a closely-spaced pair. 
This so-called hyperfine splitting  \cite{VanExter2022, Koks2022a} is observed at low q for the 1b peak, the 2b peak and the 2c peak. 
As predicted, the hyperfine splitting disappears at larger cavity lengths.

The second qualitative observation is that the fine structure at low $q$ consists of (almost) equidistant peaks. 
This shows that the anisotropic effect dominates, see Eq. \eqref{eq:DLani}. 
As $q$ increases by going to larger cavity lengths, the peaks are not equidistant anymore, suggesting that aspheric effects are taking over the lead. 
At larger cavity lengths, the resonances seem to group themselves in pairs with similar scaling. 
This is another characteristic of the aspheric contribution; since the aspheric contribution scales with $\ell^2$, mode pairs with the same $\lvert \ell \lvert$ experience similar scaling. 
The frequency splitting within these equal $\lvert \ell \lvert$ pairs results from spin-orbit coupling, which is one of the nonparaxial effects. 

\subsection{Step 4: Fine Structure (anisotropy)} \label{subsec:FineStructure_anisotropy}
The quantitative analysis of the fine structure is simple when each $N$ group is split in $N+1$ (almost) equidistant peaks, because this indicates that anisotropic effects dominate and a paraxial description suffices.  
More specifically, the ratio between the splitting $\delta (\Delta L)$ within each $N$ group over the transverse-mode spacing $\Delta L_{\rm trans}$ yields the paraxial estimate for the anistropy $\eta_{\rm ani} \approx \delta (\Delta L)/\Delta L_{\rm trans}$, see Eq. (\ref{eq:DLani}).
If we apply this equation to our data, we obtain $\eta_{\rm ani} = 0.031(3)$ for a short ($q=6$) cavity.
This estimate is based on the average $\delta \Delta L = \Delta L_{\rm fine} = 1.15(5)$~nm over two resonances at $0.89(7)$~nm and $1.20(7)$~nm, which are split due to a so-called hyperfine interaction \cite{VanExter2022}, and its comparison with $\Delta L_{\rm trans} = 37$~nm.

However, the above estimate assumes that the anisotropic effect dominates over all other effects and, therefore, tends to 
overestimate $\eta_{\rm ani}$ as follows.
In the absence of anisotropy, the nonparaxial effect is expected to split the two $N=1$ resonance by $2\Delta L_n = 0.64(4)$~nm for $R_{\rm eff} \approx 15.8(5)~\mu$m.
In the presence of anisotropy, the full theory predicts a splitting of $\Delta L_{\text{fine}}=2\sqrt{\Delta L_{n}^2 + \Delta L_{a}^2}$. 
Substitution of the measured $\Delta L_{\rm fine} = 1.15(5)$~nm and $2\Delta L_n = 0.64(4)$~nm yields $2\Delta L_a = 0.95(8)$~nm and a new estimate of $\eta_{\rm ani} = 0.026(3)$. 
We have repeated this complete analysis for all $N=1$ pairs in the range of $5\leq q \leq 20$. 
This extended analysis yields the more reliable estimate $\eta_{\text{ani}}=0.024(3)$.

Whether a simple paraxial analysis suffices or not should generally be visible in the fine structure of the modes in the $N \geq 2$ groups.
If anisotropy dominates (top corner in Fig. 1), each group is expected to comprise $N+1$ modes with an equidistant spacing that equals the splitting of the $N=1$ pair.
But if the spacings are different, other effects must play a role.  
For cavities with large $R_m$, and hence weak nonparaxial effects, the aspheric mirror shape might be the culprit. 
For cavities with small $R_m$, nonparaxial effects will probably play a role. 
The cavities studied in ref. \cite{Koks2022b} were exceptional because they were almost rotationally symmetric, with anistropies as small as $\eta_{\rm ani} \approx 0.006$, and also had relatively small $R_m$. 
They thus operated in the regime were nonparaxial effects dominate (bottom-right corner in Fig. 1), and where vector LG modes, instead of HG modes, are the eigenmodes.

\subsection{Step 5: Fine Structure (complete)} \label{subsec:FineStructure_complete}

The analysis presented above provided estimates for all three mirror parameters. 
From the Gouy phase $\chi_0(L)$, $R_m$ and a rough estimate of $h$ have been determined.
A comparison between cavity and planar modes provided a more accurate estimate of $h$. The analysis of the fine structure of the $N=1$ group gave an estimate of $\eta_{\text{ani}}$. 
A more detailed analysis of the fine structure of the higher-order modes can provide additional data to check and improve these estimates. 
We will perform this analysis for the $N=2$ fine structure. 

From the matrix in Eq. \eqref{eq:N2} one can compute the eigenvalues. 
As these analytical expressions become impractical, we will resort to a computational approach. 
A $\chi^2(R_m, h,\ea)$ analysis of the observed $N=2$ fine structure over a range of $5\leq q \leq 18$, yields  $\ea=0.0249(3)$,  $R_m=18.4(2)\ \mu$m and $h_4=0.324(5)\ \mu$m, where the errors are exclusively statistical. 
Although the estimate for $\ea$ is consistent with prior estimates, $R$ and $h$ deviate from this.
The origin of the discrepancy in the radius is suspected to lie in the limitation of using $R_m$ as a parameter for both the mirror curvature and the wavefront curvature, i.e. in the relation $R_{\rm eff} \neq R_m$. 
The discrepancy in $h$ arises from shape-deviations from a perfectly Gaussian shaped mirror. 
Both discrepancies are discussed in Sec. \ref{sec:comparison}.
Despite these discrepancies, the competition between the various mirror-shape and non-paraxial effects has been observed and explained both qualitatively and quantitative.

\subsection{Finesse}
\label{sec:Finesse}

\begin{figure}
    \centering
    \includegraphics[width=7.9cm]{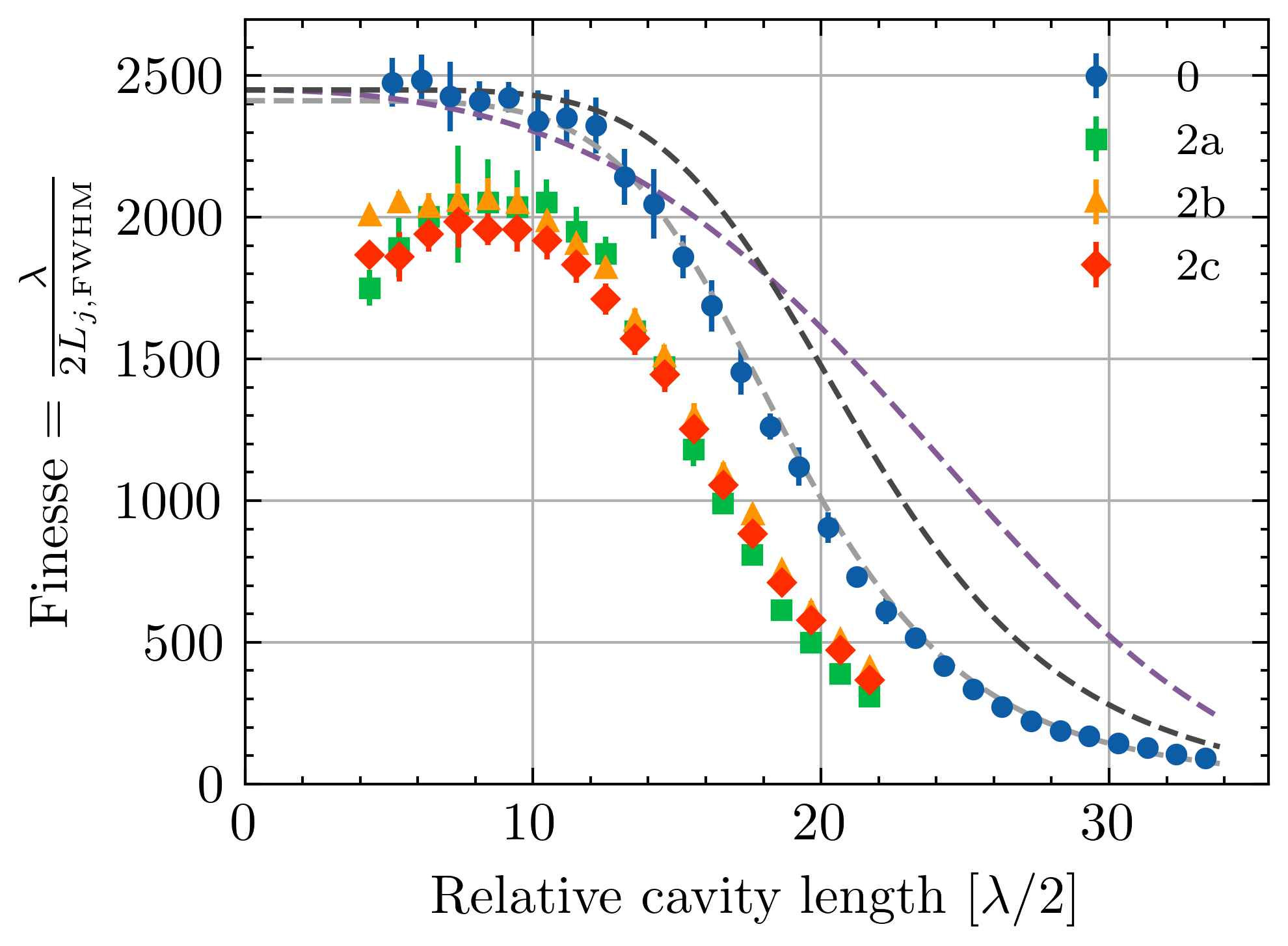}
    \caption{Finesse versus cavity length for the fundamental mode (blue) and the $N=2$ modes (green, orange, red). The gray dashed curve provides a `guide to the eye'. The purple dashed curve is the finesse predicted by scattering losses according to Eq. \eqref{eq:loss1}. The black dashed curve is the finesse predicted by deflection losses according to Eq. \eqref{eq:loss2}.}
    \label{fig:finesse2}
\end{figure}

Figure \ref{fig:finesse2} shows the measured finesse, $F_j = L_{\rm FSR}/\delta L_{j,{\rm FWHM}} \approx \lambda/2\delta L_{j,{\rm FWHM}}$ as a function of cavity length, with $\delta L_{j,{\rm FWHM}}$ the full width at half maximum of the resonance peaks. The blue data points present the fundamental mode, while the other colors represent the three $N=2$ modes.
For the fundamental mode, the finesse is approximately constant at $F \approx 2400 \pm 100$ for short cavities ($q \leq 12$) and then rapidly decreases for longer cavities.
A `guide to the eye' for the fundamental mode is provided by a function of the form $F = F_0 / (1+ \alpha L^6)$ presented by the gray dashed curve.
This function, which is based on a Taylor expansion of Eq. \eqref{eq:loss2} around $L=R_m/2$ for $hk = 6$, provides a surprisingly nice fit.
For many $N=2$ modes, the finesse at short cavity lengths ($q\leq 12$) is complicated to determine due to hyperfine splitting. 
Finesse data at longer cavity lengths ($q > 21$) for the $N=2$ modes are absent as higher-order modes could not be efficiently excited.
The purple and black dashed curves in Fig. \ref{fig:finesse2} show the predicted finesse for scattering losses, Eq. \eqref{eq:loss1}, and deflection losses, Eq. \eqref{eq:loss2}, respectively, for $R_m=13.6$ and $h=0.61$. 
Although neither the scattering nor the deflection loss models reproduce the data quantitatively, the finesse drop is qualitatively best reproduced by the somewhat speculative equation for deflection losses. 
This suggests that deflection loss, as a generalization from clipping loss, might indeed be the underlying physics to the finesse drop.

In general, larger cavities suffer from clipping and mode-coupling loss \cite{Benedikter2015,Benedikter2019,Pallmann2023}, which often 
limits their operation regime to $L < R_m/2$. 
The finesse in our cavity already drops at shorter length, namely at $q \approx 14$ instead of the $q \approx 21$ calculated for $L=R_m/2$, presumably because the mirror depth $h$ is low. 
To estimate this Gaussian-mirror-shape effect, we compare with simulations performed by Benedikter et al. \cite{Benedikter2015}. 
The simulations corresponding to our geometry, with $\epsilon=\sqrt{2\eta/(1+\eta)}=0.24-0.26$ and $\omega_c / a = \sqrt{\lambda/(2\pi h)}\approx 0.56$ in their notation, predict a rapid decrease in finesse already at $L=R_m/7$. 
In our cavity, this corresponds to $q=6$ which is twice as short as the $q = 12$ value that we observe.

We thus attribute the observed leakage at large cavity length to the combination of the shallow depth $h$ of our concave mirror, which makes the cavity more susceptible to leakage, with the non-spherical shape of the mirror, which requires admixture of high-order modes to match the modal wavefront to the mirror surface. 
We observe this leakage in two different ways. 
In direct images of the cavity, we observe this leakage as an intensity pattern outside the cavity that has an intriguing ring-shaped structure and decays approximately as the inverse of the distance to the cavity.
In far-field/angular images, we observe this leakage as Fresnel rings with an angular spacing that agrees with the one expected from the distance between the planar parts of both mirrors.

\subsection{Polarization tomography}
\label{sec:Tomography}

\begin{figure}
    \centering
    \includegraphics[width=7.9cm]{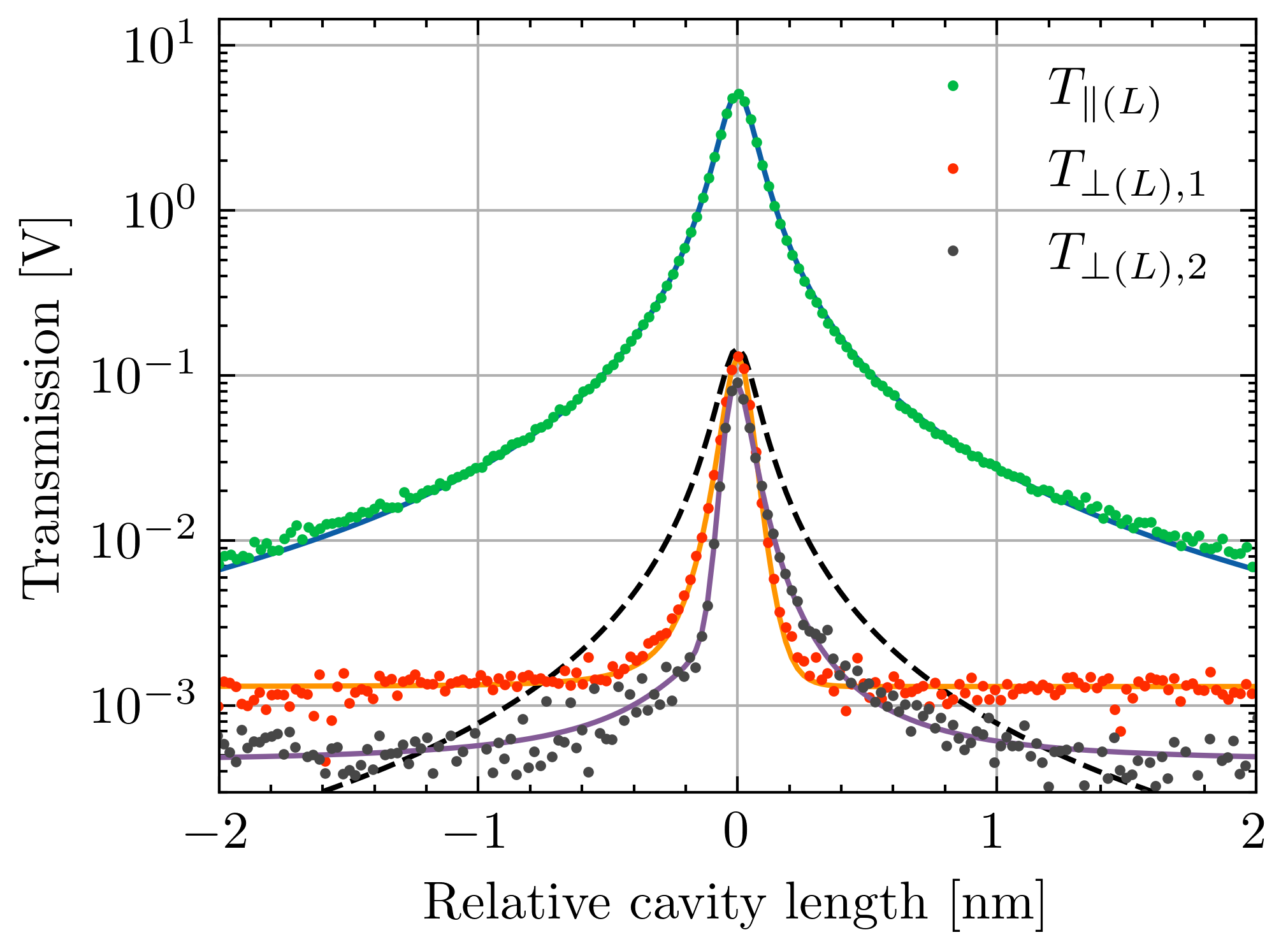}
    \caption{Transmission spectrum of the fundamental mode, for parallel polarization (green) and two crossed polarization (orange and purple for $\theta_{\rm in} = \pm 45^\circ$). The black dashed curve is a scaling of the green/blue Lorentzian and serves as a reference for comparison for crossed polarization spectra.}
    \label{fig:PolarizationTomography}
\end{figure}

The $T(L)$ transmission spectra presented above were measured without polarization selection. 
Additional information can be obtained when the transmission is measured behind a polarizer for various settings of the input polarization angle $\theta_{\rm in}$ and the output angle $\theta_{\rm out}$.
Figure \ref{fig:PolarizationTomography} shows typical results obtained for the fundamental $N=0$ mode(s).
The signal observed for parallel polarization ($\theta_{\rm out} = \theta_{\rm in}$) is typically two orders of magnitude stronger than the signal observed for crossed polarization.
This shows that the transmission for $N=0$ modes copies the input polarization to a large extent, but not completely.
Furthermore, the polarization contrast $T_\perp(L)/T_\parallel(L)$ depends on the input polarization. 
If the input polarization is aligned with the axes of the elliptical mirror, 
the polarization contrast is very high, presumably because the input then only excites one of the two fundamental eigen modes.
The $T_\perp(L)$ spectrum is strongest when the input polarization is rotated by $\pm 45^\circ$ with respect to this setting.
The two $T_\perp(L)$ curves in Fig. \ref{fig:PolarizationTomography} were measured under these conditions.
The fit curves show that these $T_\perp(L)$ spectra do not have a Lorentzian shape, as $T_\parallel(L)$ has, but are more narrow and have the expected ``double-Lorentzian" shape described in Appendix \ref{sec:appendix-Tomography}, with some asymmetry due to $\theta_{\rm out} - \theta_{\rm in} \neq 90^\circ$. 
The observed ratio $T_\perp(L)/T_\parallel(L) \approx 1/50$ allows one to determine the shift between the two (orthogonally-polarized) $N=0$ resonances to be $\Delta L_{HV}/2\delta L \approx 1/\sqrt{50} \approx 0.14$, even though these resonances cannot be resolved individually. 
This is comparable to the value expected from the theory described in Appendix \ref{sec:appendix-Tomography}, because 
\begin{equation}
    \frac{\Delta L_{HV}}{2 \delta L} = \ea \frac{F\lambda}{2\pi^2 R_m} \approx 0.11\,,
\end{equation}
for our cavity with $\ea \approx 0.025$, $F \approx 2500$ and $R_m \approx 18$~$\mu$m. 

For $N \geq 1$ modes, we observe much stronger $T_\perp(L)$ signals at a typical level of 20-50 \% of the $T_\parallel(L)$ signals. 
This shows that these modes do not simply copy the polarization of the input beam but have a strong polarization preference of their own and that this polarization preference is in the form of a polarization pattern.
We propose to define the ``vectorness" of this pattern/mode via the minimum contrast ratio $T_\perp/T_\parallel$ observed with crossed polarizers. 
With this definition, which is similar to the definition of vectorness used in Stokes polarimetry \cite{Singh2020}, the $N=0$ modes have almost no vectorness while the $N \geq 1$ have a typically vectorness of $0.2-0.5$.
The mere observation that resonant modes have vectorness, i.e. that the optical transmission is not zero between crossed polarizers, shows that 
their vector fields $\vec{E}_j(x,y) \neq \vec{e}_j E_j(x,y)$. 
The amount of vectorness can be used to quantify the strength of the spin-orbit or $\vec{L} \cdot \vec{S}$ coupling \cite{Bliokh2015}, relative to the scalar mode shaping effects.
The vectorness of the $N \geq 1$ modes might actually be the most prominent and easily observable nonparaxial effect.

\subsection{Reflection Spectrum}

\begin{figure}[ht]
    \centering
    \includegraphics[width=7.9cm]{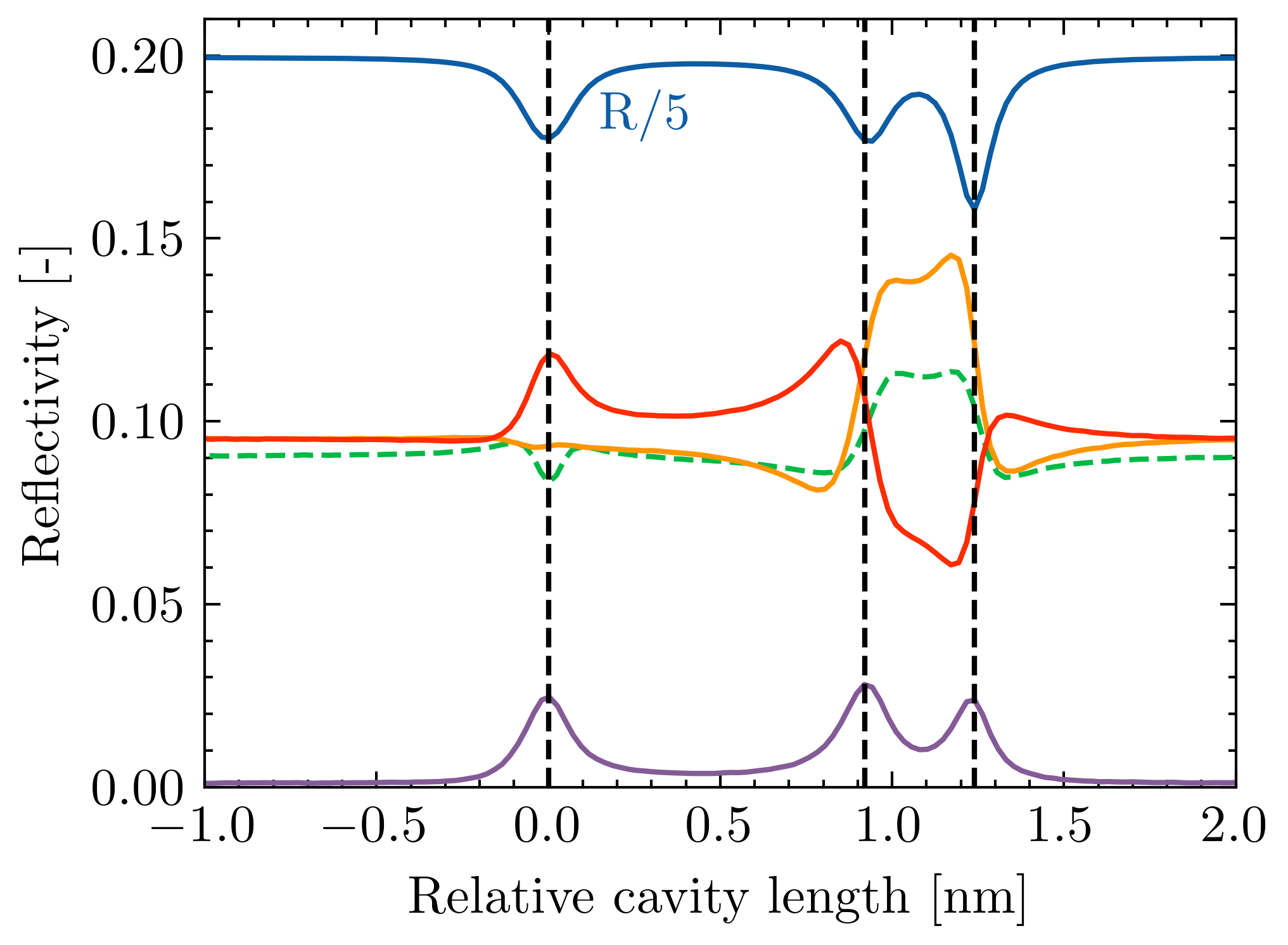}
    \caption{Reflectivity versus relative cavity length for the $N=1$ group, under different polarization projections. The top/bottom curves are reflection spectra observed under parallel/perpendicular polarization, relative to the input. The three middle curves show spectra at polarization angles of $-18^\circ$ (red) and $+18^\circ$ (orange and dashed green at different spatial alignment) relative to perpendicular, where the off-resonant reflectivity $R \approx \sin^2(18^\circ) \approx 0.095$.}
    \label{fig:reflection-spectrum}
\end{figure}

Figure \ref{fig:reflection-spectrum} shows the reflectivity versus relative cavity length for the $N=1$ group, under various polarization conditions. 
The reflection dips observed under parallel polarization ($R/5$ in top curve to reduce 100\% reflection to a value of 0.200) become reflection peaks under perpendicular polarization. 
The relative strength of these peaks ($R \approx 0.02$) contains information on the depolarizing nature (or vectorness) of these modes and their mismatch with the input polarization.  
The three middle curves show polarization-resolved spectra at $\pm 18^\circ$ relative to perpendicular. 
The resonances in these spectra are more intriguing. 
They combine absorption and dispersion features, in so-called Fano profiles \cite{Fano1961}, and show interference between neighboring resonances, an interference that changes sign when the analyzing polarizer is rotated from $+ 18^\circ$ to $- 18^\circ$.
We have observed similar features for resonances in the $N=2$ group and in different microcavities. 

The measurements depicted in Fig. \ref{fig:reflection-spectrum} were taken under misaligned conditions, to divide the input power over many transverse modes. 
When the input is aligned with the fundamental mode, such that this mode receives about $90~\%$ of the input power, we observe a dip of $\approx 70~\%$ and a peak transmission of $\approx 60~\%$ at resonance.
These numbers show that the scattering losses in our mirrors are much smaller than the transmission losses.
This is in agreement with the finesse observed in short cavities, which is only slightly below the value calculated from the mirror losses only. 


\section{Resonant mode profiles}
\label{sec:profiles}

\subsection{Qualitative Observations} 
\label{subsec:mode_profile_qualitative}

To measure the resonant mode profiles, we slowly scan the cavity length over one FSR, typically in 50 seconds. 
A movie of the far-field intensity transmission profile is shot with a polarization-resolving camera. 
By summing all pixel values to determine the power in each frame, this movie can be converted into a transmission spectrum. 
Accurate length calibration is unreliable for such a slow scan due to mechanical vibrations. 
However, the chronological order of peaks clearly represents the cavity spectrum. 
The resemblance with the transmission spectra measured with the PMT, such as in Fig. \ref{fig:transmission spectrum}, allows one to match the resonances and label them accordingly. 

\begin{figure*}
    \centering
    \includegraphics[width=\linewidth]{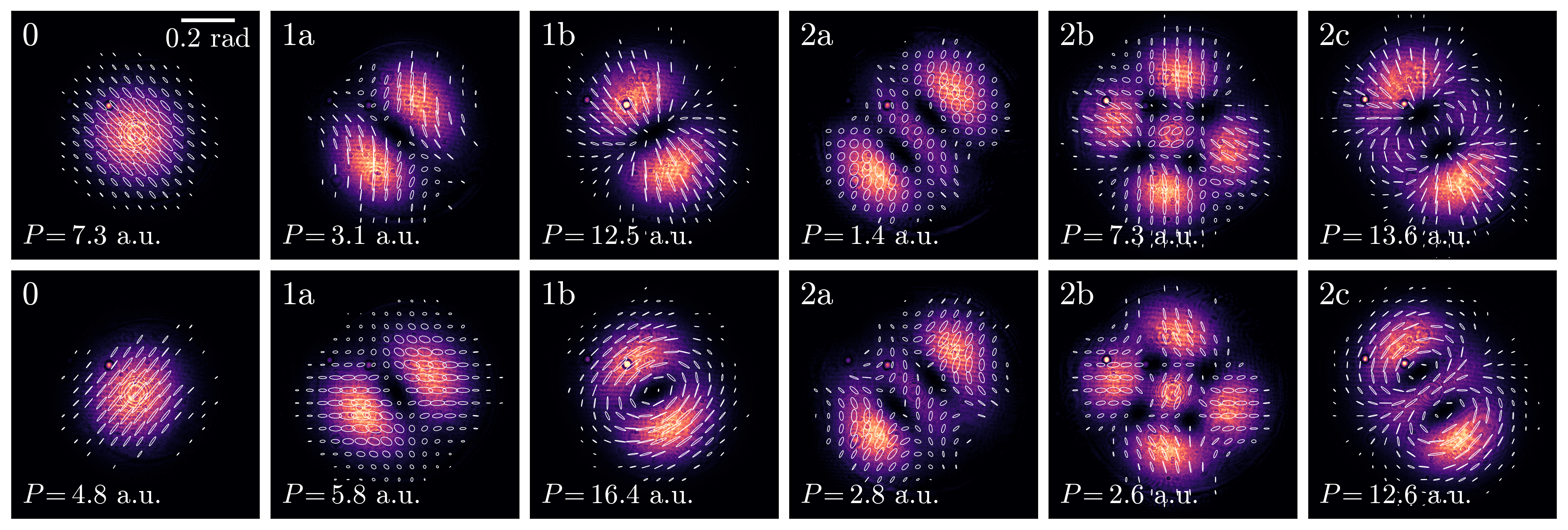}
    \caption{Mode profiles of the $N = 0$, $1$ and $2$ modes in the lab frame, for different laser polarizations: anti-diagonal (top row) and diagonal (bottom row). The fundamental mode copies the input polarization.
    All measurements have been taken with anti-diagonal misalignment. For the $N=1$ and $2$ modes, the input polarization is reproduced along the misaligned direction of the mode profiles. The integrated power of the image is presented in the bottom left in arbitrary units.
    Each polarization-resolved image has been normalized individually.
    }
    \label{fig:modes_pol_and_alignment}
\end{figure*}

The mode profiles presented below have mainly been measured at the shortest cavity length, in the $q=5$ group. 
To better observe the high-order modes, the setup has been purposely misaligned, by displacing the laser in various directions. 
To study the polarization profiles of the modes, the polarization of the input laser has also been varied.

Figure \ref{fig:modes_pol_and_alignment} shows the combined effects of misalignment and polarization on the resonant $N = 0$, $1$ and $2$ modes. 
The polarization profiles are visualized by polarization ellipses (representing the local polarization) on top of the intensity false-color plot
The top row shows the mode profiles for anti-diagonal input polarization and the bottom row for diagonal input polarization. 

At first sight, the intensity profiles resemble HG modes. 
However, further inspection shows deviations from the HG modes. 
For example, the non-zero intensity in the center of the 2b mode is not present in the pure HG$_{11}$ mode. 
Furthermore, the vortex- and anti-vortex-like polarization profiles \cite{Nasalski2014} are a characteristic of the vector LG modes. 
This \textit{vectorness} in the profiles is a signature of spin-orbit coupling.

As we have also seen in the transmission spectra, different alignments will change the amplitude of particular modes. 
In particular, in Fig. \ref{fig:modes_pol_and_alignment} the laser was spatially misaligned along the anti-diagonal direction in the lab frame. In the $N=1$ group, the 1b mode has the highest intensity, and similarly the 2c mode had the higher intensity in the $N=2$ group. This can be explained by their profile being elongated along the misaligned direction. Similar results have been observed for diagonal misalignment (figure not shown), where the 1a and 2a modes then contain the most power in their $N$-group.

The polarization profiles of the $N=1$ group resemble the vector LG polarization profiles \cite{Yu1983}. 
The 1a mode resembles the 1B profiles, while the 1b mode resembles the 1A profiles.
The $N=0$ mode is excited in the same polarization as the input polarization. The $N\geq 1$ modes copy the input polarization along the misaligned direction. The observed effects of alignment and polarization are in line with theory, as one expects to excite the modes which have the largest overlap with the inserted laser profile. Similar results, consistent with this prediction, have been obtained for diagonal misalignment, with various input polarizations.

\subsection{Modal Decomposition} \label{subsec:mode_profile_recipe}
\begin{figure}
    \centering
    \includegraphics[width=7.9cm]{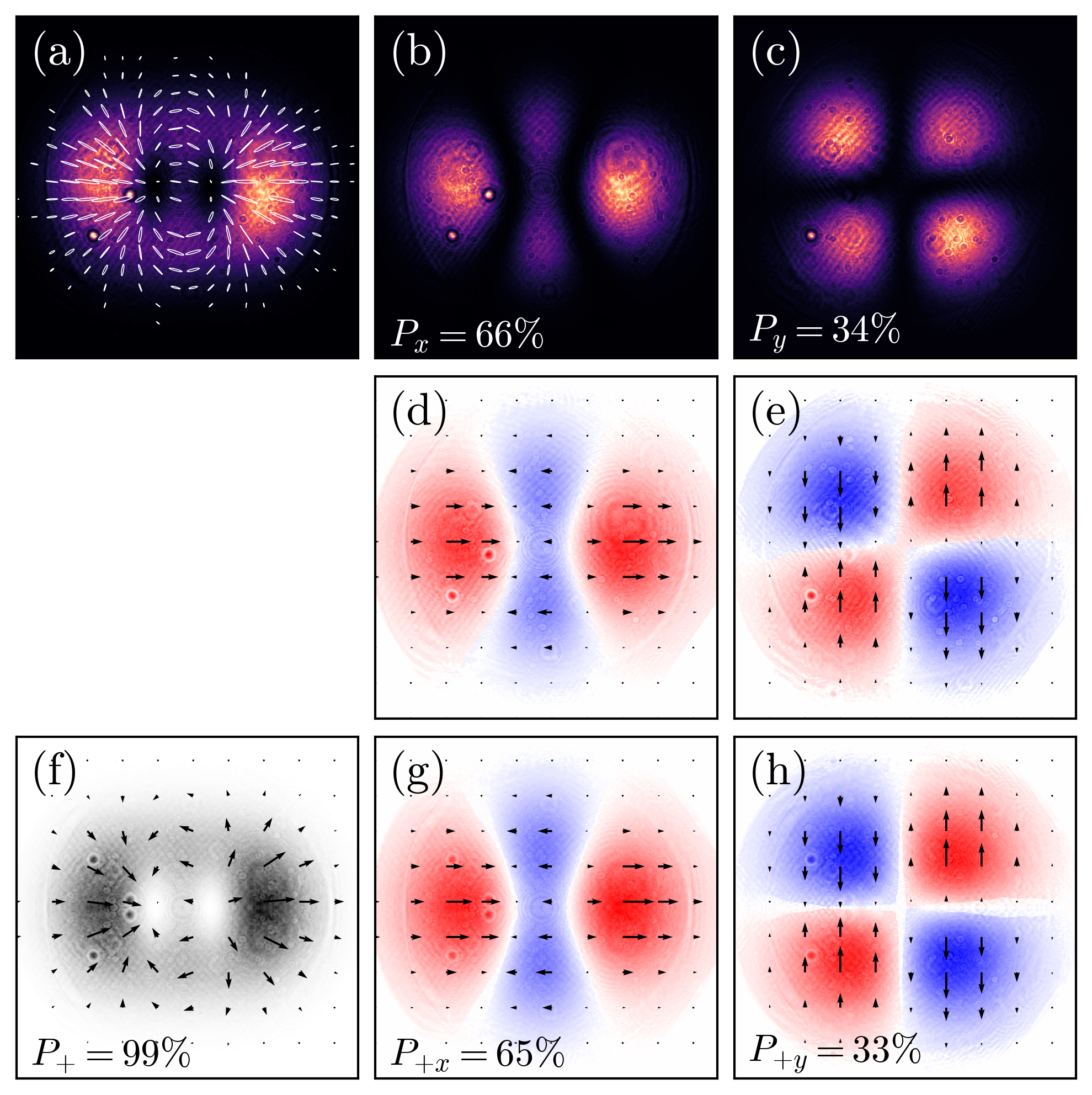}
    \caption{(a) Mode profile $I(x,y)$ of the 2a mode in the mirror frame, with polarization profiles represented by overlying polarization ellipse. The measurements are taken with  both vertical misalignment and vertical input polarization with respect to the mirror frame. (b-c) Splitting into $I_x (x,y)$ and the $I_y (x, y)$. (d-e) The extracted $E$-field for the $x$ and $y$ polarizations. (f-h) The $-$ symmetric part of the $E$-field: (f) the vector field, (g) x-polarized part, (h) y-polarized part.}
    \label{fig:modes_recipe}
\end{figure}

The mode profiles will be analyzed to find their modal decomposition in the basis of HG-modes. 
The reflection-symmetric axes of the observed mode profiles are rotated with respect to the \textit{lab frame}: the frame of the setup. 
These symmetry axes are presumably caused by the direction of the mirror anisotropy. 
The frame set by these 
axes will therefore be referred to as the \textit{mirror frame}. 
As a first step in the analyses, 
the images are rotated into the mirror frame. 
The perceived rotation angle can vary for different frames by $\pm 5^{\circ}$, due to mode mixing.
We estimate the mirror axis to be at a polar angle of $-55^{\circ} \pm 5^{\circ}$ and $35^{\circ} \pm 5^{\circ}$ in the lab frame. 
For the remainder of this section, all images are rotated accordingly and will be presented in the mirror frame.
The complete analysis will be presented for the 2c mode profile shown in Fig. \ref{fig:modes_recipe} (a), which is the rotated version of the 2c mode presented in the top row in Fig. \ref{fig:modes_pol_and_alignment}. 


After rotation, the next step is to find the $\vec{E}$-field to decompose the image into even and odd `+/-' modes, as discussed in section \ref{subsec:symmetry}. 
First $I_x (x,y)$ and $I_y (x, y)$ are separated, as shown in Fig. \ref{fig:modes_recipe} (b-c). 
One already recognizes the HG$_{20}$ and HG$_{11}$ modes. 
Then one takes the square root and finally one assigns plus and minus signs.
Assigning plus/minus signs to regions, separated by zero-intensity boundaries, should be done in such a way that neighboring regions have opposite signs. This way the signs can be determined up to an overall sign.
The determined $E_x$-field and $E_y$-field for the 2c mode are presented in Fig. \ref{fig:modes_recipe} (d) and (e) respectively.
Now that the $\vec{E}$-field has been determined, we can also separate the symmetric and antisymmetric parts as described by Eq. \eqref{eq:split_symm_Efield}. 
For this polarization and alignment, the 2c mode is dominantly $+$ symmetric ($99\%$), as shown \ref{fig:modes_recipe} (f-h). 
The $-$ symmetric part only contains some residuals ($3\%$, not shown).

The third and final step is to calculate the overlap integrals.
We find that $\braket{ E_{+,x}}{HG_{20}} = 0.92$, $\braket{ E_{+,x}}{HG_{02}} = -0.27$ and $\braket{ E_{+,y}}{HG_{11}} = 0.97$. After inclusion of the measured power ratio ($66\%$ vs $34\%$ for $P_x$ and $P_y$ resp.), the resonant mode can thus be written as
\begin{multline} \label{eq:2a_decomposition}
    \vec{E}_{2a,-} \approx  0.75\times HG_{20} \vec{e}_x + 0.57\times HG_{11} \vec{e}_y \\-0.21\times HG_{02} \vec{e}_x\,,
\end{multline}
where $\vec{E}_{j} = \vec{E}_{j}(x,y)$ and $HG_{mn}=HG_{mn}(x,y)$. This mathematical expression has an overlap of $\sqrt{0.75^2 + 0.57^2 + 0.21^2} = 0.97$ with the measured mode.
We thus interpret the 2c as a HG$_{20}\vec{e}_x$ mode with mixing of the HG$_{11}\vec{e}_y$ mode due to spin-orbit coupling, which is one of the nonparaxial effects. 
The mixing of the HG$_{02}$ mode into the HG$_{20}$ mode is small. 
We thus conclude that the anisotropic contribution is dominant over the aspheric contribution $X > (1-G)$ and the HG dynamical matrix in Eq. $\eqref{eq:N2_HG}$ is dominated by its diagonal and next-to-diagonal elements.

This conclusion agrees with the findings in Sec. \ref{sec:spectra}. More specifically, from the parameters estimated in that section for $N=2$ ($\ea=0.0249(3)$,  $R_m=18.4(2)\ \mu$m and $h=0.324(5)\ \mu$m), the 2c symmetric eigen mode is predicted to be $(0.86, 0.47, -0.19)$. Computation of the overlap integral of the predicted mode with the measured mode yields an overlap of $0.95$.

\begin{figure}
    \centering
    \includegraphics[width=7.9cm]{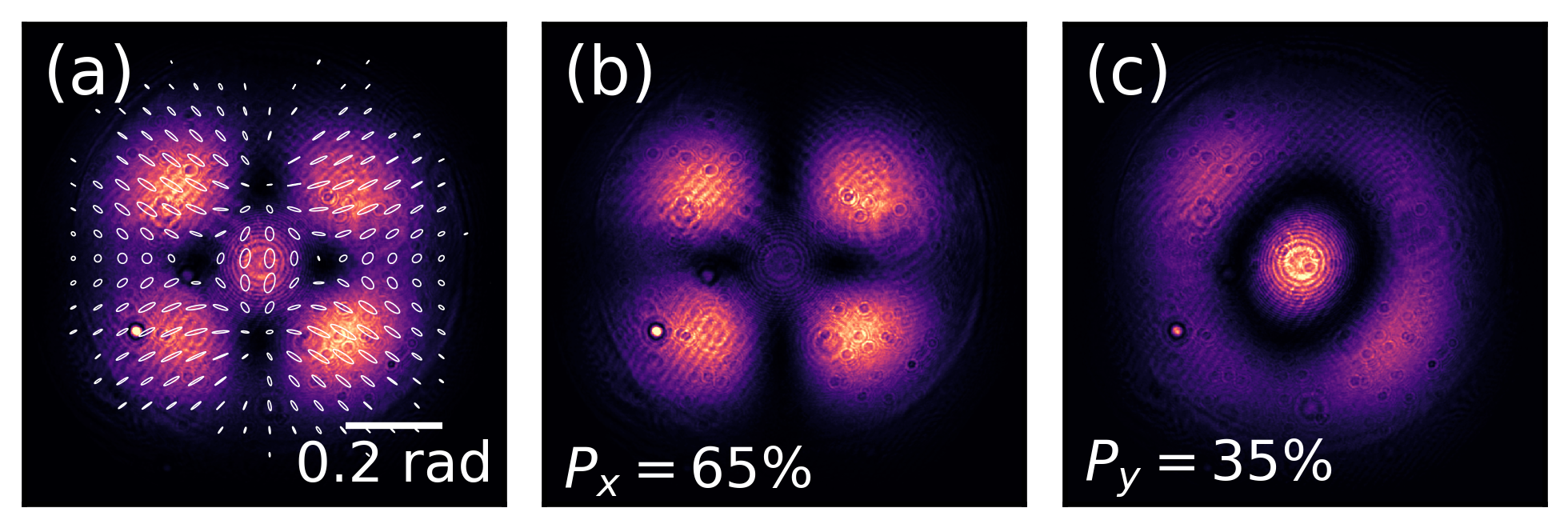}
    \caption{(a) The $I(x,y)$ mode profile of the 2b mode in the mirror frame. (b) The $I_x(x,y)$ profile resembles a HG$_{11}$ mode. (c) The $I_y(x,y)$ profile resembles a LG$_{10}$ mode.}
    \label{fig:modal_decomposition_2b}
\end{figure}

We have applied the above three-step analysis also for the 2b mode presented in the top row in Fig. \ref{fig:modes_pol_and_alignment}.
In the $I_x (x,y)$ and $I_y (x, y)$ profiles shown in Fig. \ref{fig:modal_decomposition_2b}, one recognizes the HG$_{11} \vec{e}_x$ and LG$_{01} \vec{e}_y$ modes, respectively. After decomposition, we find that the profile is $98\%$ $-$ symmetric and can be expressed as

\begin{multline} \label{eq:2b_decomposition}
    \vec{E}_{2b,+}\approx +0.43 \times HG_{20} \vec{e}_y + 0.77\times HG_{11} \vec{e}_x \\+ 0.37\times HG_{02} \vec{e}_y \,.
\end{multline}
Although the HG$_{11}$ is already recognized in the $I(x,y)$ profile in Fig. \ref{fig:modal_decomposition_2b}, we can now also understand the non-zero center, where the field is $x$- instead of $y$-polarized. 
The mode can be understood from theory, where Eq. \eqref{eq:N2_HG} shows that the HG$_{11}$ mode receives contributions from the HG$_{02}$ and HG$_{20}$ mode due to spin-orbit coupling.
The $-$ symmetric eigen mode predicted by the characteristic parameters is $(0.51 , 0.82 , 0.26)$. 
The predicted mode has an overlap with the image of $0.94$. 

For the weakly excited 2a modes in Fig. \ref{fig:modes_pol_and_alignment}, we find that both the $+$ and $-$ symmetries are present in equal strength ($50\pm3\%$). 
For a different alignment that excited the 2a modes stronger, one of the two polarization profiles and symmetries could be dominantly excited with the proper input polarization. 
In that alignment, the 2c mode was only weakly excited, but both polarization profiles could still be observed separately because of the hyperfine splitting of the 2c mode.

The recipe has been repeated with different input conditions for all $N=2$ modes. 
All modes had a dominant symmetry ($\geq84\%$) except for the two 2a modes presented in Fig. \ref{fig:modes_pol_and_alignment}. 
We note that the strength of the dominant symmetry strongly depends on the perceived rotation of the mirror frame. 
By allowing a free rotation axis per image in the analysis, we find that the presence of the dominant symmetry increases from $>84\%$ to $>89\%$.
We conclude that all the measured modes in the mirror frame of $-55^{\circ}$ w.r.t the lab frame, with a dominant symmetry, have an overlap $>0.87$ with the predicted mode of the dominant symmetry. More specifically their dominantly symmetric parts have an overlap of $>0.95$ with the predicted mode of that symmetry.
The effects of mode mixing in the mode profile is well explained by the theory and the estimated parameters.

\subsection{Aspheric effects}
\begin{figure}
    \centering
    \includegraphics[width=7.9cm]{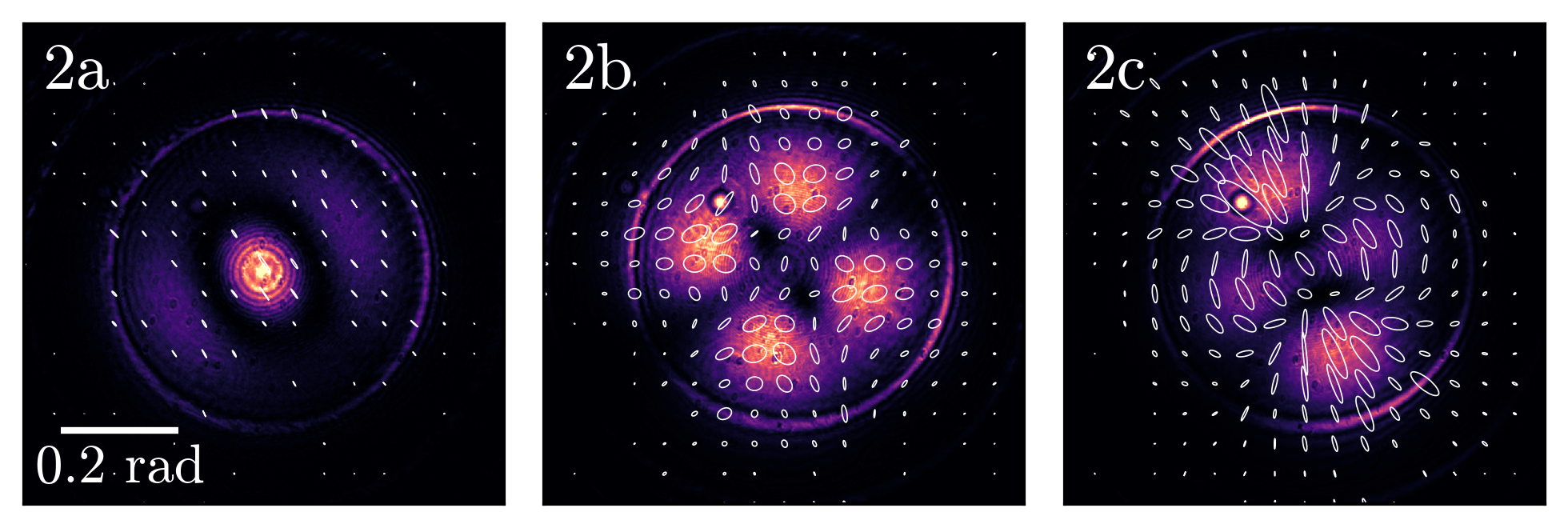}
    \caption{The $N=2$ modes at $q = 18 \pm 1$, in the lab frame. Losses are observable in the outer region of the profile. The $2a$ modes is linearly polarized, thus $\ell=0$. The polarization pattern of the $2b$ and $2c$ modes resemble those expected for the $\ell=2$ LG vector modes.}
    \label{fig:N2_highZ}
\end{figure}

Fig \ref{fig:N2_highZ} presents the mode profiles of the $N=2$ modes at larger cavity lengths, $q = 18 \pm 1$ in the lab frame. 
These mode profiles clearly deviate from the $N=2$ modes observed in figure \ref{fig:modes_pol_and_alignment}. 
The ring-shaped features in these (angle-resolved) figures originate from the leakage of the modes discussed in Sec. \ref{sec:mirror-shape}.
That the mode profiles have a smaller opening angle than those at shorter cavity lengths is due to their larger waist, since $\theta_0 = \lambda/\pi w_0$.

The third observation is that these mode profiles closely resemble LG vector modes. 
The $2a$ mode has a non-zero radial mode number $p > 0$ and a linear polarization, hence $\ell=0$. 
The 2b and 2c mode profiles closely resemble the LG 2B and 2A polarization profiles respectively, where $\ell=2$. 
This observation agrees with the conclusion drawn in Sec. \ref{subsec:FineStructure}, where Fig. \ref{fig:fine_structure_peaks} showed pairing of modes at increasing $q$, suggesting strong aspheric $\ell^2$ effects.
Similarly, the mode profiles of the $4d$ and $4e$ (not shown) resemble the LG 4B and 4A modes, where $\ell=4$. 
The other observed $N=4$ modes were not strong enough to draw analogous conclusions.

\subsection{Profiles of $N\geq 3$ modes}

\begin{figure}
    \centering
    \includegraphics[width=7.9cm]{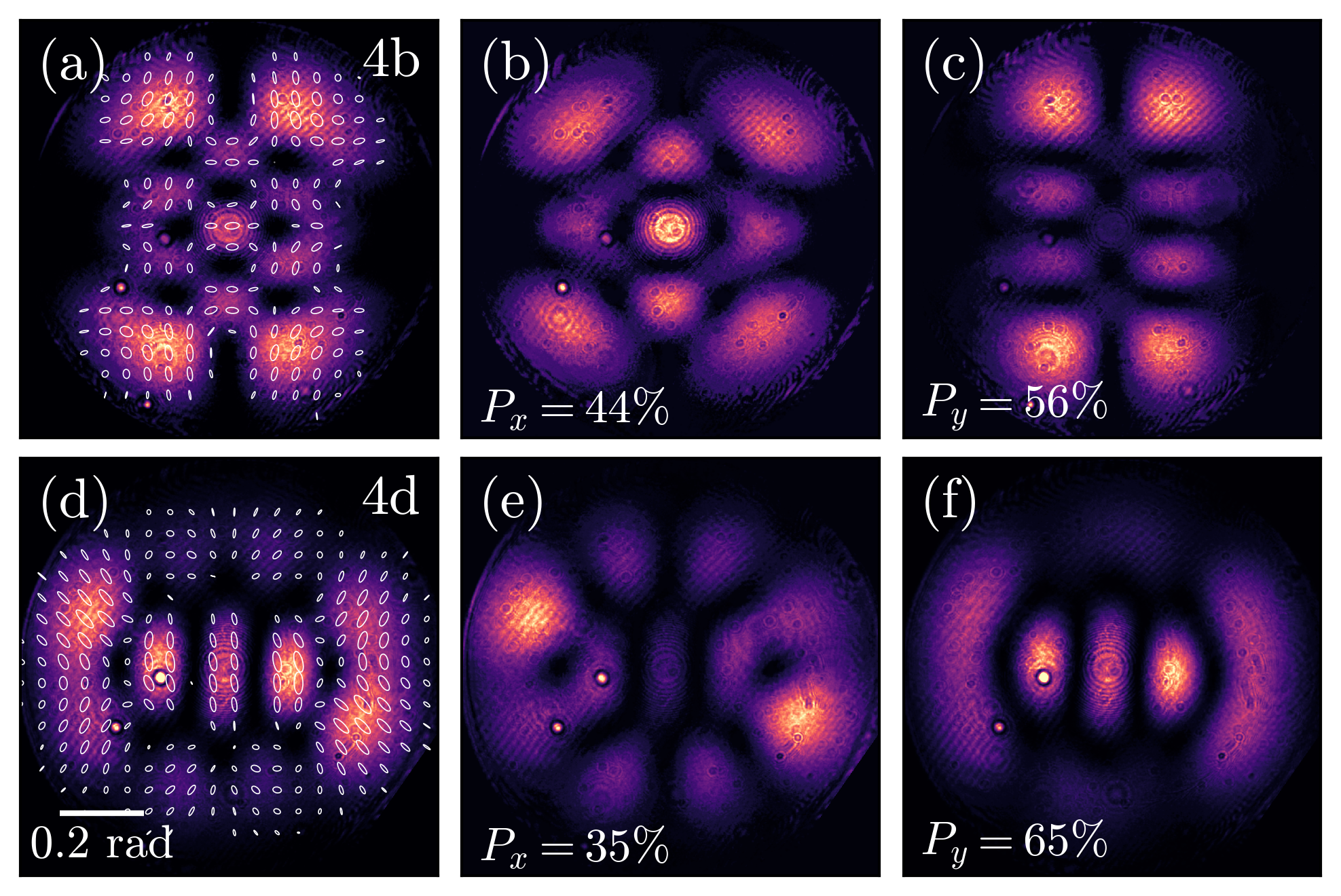}
    \caption{Decomposition of 4b mode (top row) and 4d mode (bottom row):
    (a) the $I(x,y)$ mode profile of the 4b mode in the mirror frame,
    (b) its $I_x(x,y)$ profile can be understood as a mix of the HG$_{04}$, HG$_{22}$, and HG$_{40}$ modes,
    (c) its $I_y(x,y)$ profiles resemble a HG$_{13}$ mode,
    (d) mode profile of the 4d mode,
    (e) its $I_x(x,y)$ profile is a mixture of modes,
    (f) its $I_y(x,y)$ profile is dominated by the a HG$_{04}$ mode.
    }
    \label{fig:modal_decomposition_4b_4d}
\end{figure}

We have repeated the modal analysis for several high-order modes and checked for consistency. 
Figure \ref{fig:modal_decomposition_4b_4d} shows the profile of the 4b (top row) and 4d (bottom row) modes of the same measurement series as in the bottom row of Fig. \ref{fig:modes_pol_and_alignment}.
The 4b mode is $97\%$ $+$ symmetric and the decomposition of the mode reveals it's mode admixture as 
\begin{multline} \label{eq:decomposition}
    \vec{E}_{4b,+}\approx +0.11 \times HG_{40} \vec{e}_x + 0.58\times HG_{22} \vec{e}_x \\ -0.72\times HG_{13} \vec{e}_y + 0.25\times HG_{04} \vec{e}_x \,,
\end{multline}
where we omitted the weak $HG_{31} \vec{e}_y$ term. The $y$-component is dominated by the HG$_{13}$ mode, with neglectable contribution of the  HG$_{31}$.
The $x$-component is not trivial to interpret by eye, but the decomposition resolves that it can be decomposed into the HG$_{22}$, with contributions from both HG$_{04}$ and HG$_{40}$. This mixing is a result of the $(1-G)$ term in the dynamical matrix.

The $y$-component of the 4d mode can be understood as the HG$_{04}$ with small admixture of the HG$_{22}$ and an even smaller contribution of the HG$_{04}$ modes. The $x$-component however is not easily understood, due to the absence of clear zero-intensity boundaries.

Not all high-order modes are accessible for decomposition. Some images, like the $x$-component of the 4d mode, did not show clear zero-boundaries in their intensity profile, hindering the reconstruction of the $\vec{E}$-field. The absence of zero-boundaries might be a result of leakage/loss of the mode combined with weak intensity profiles and/or the absence of a dominant symmetry in the profile.

\section{Comparison of mirror profiles}
\label{sec:comparison}

\begin{figure}
    \centering
    \includegraphics[width=7.9cm]{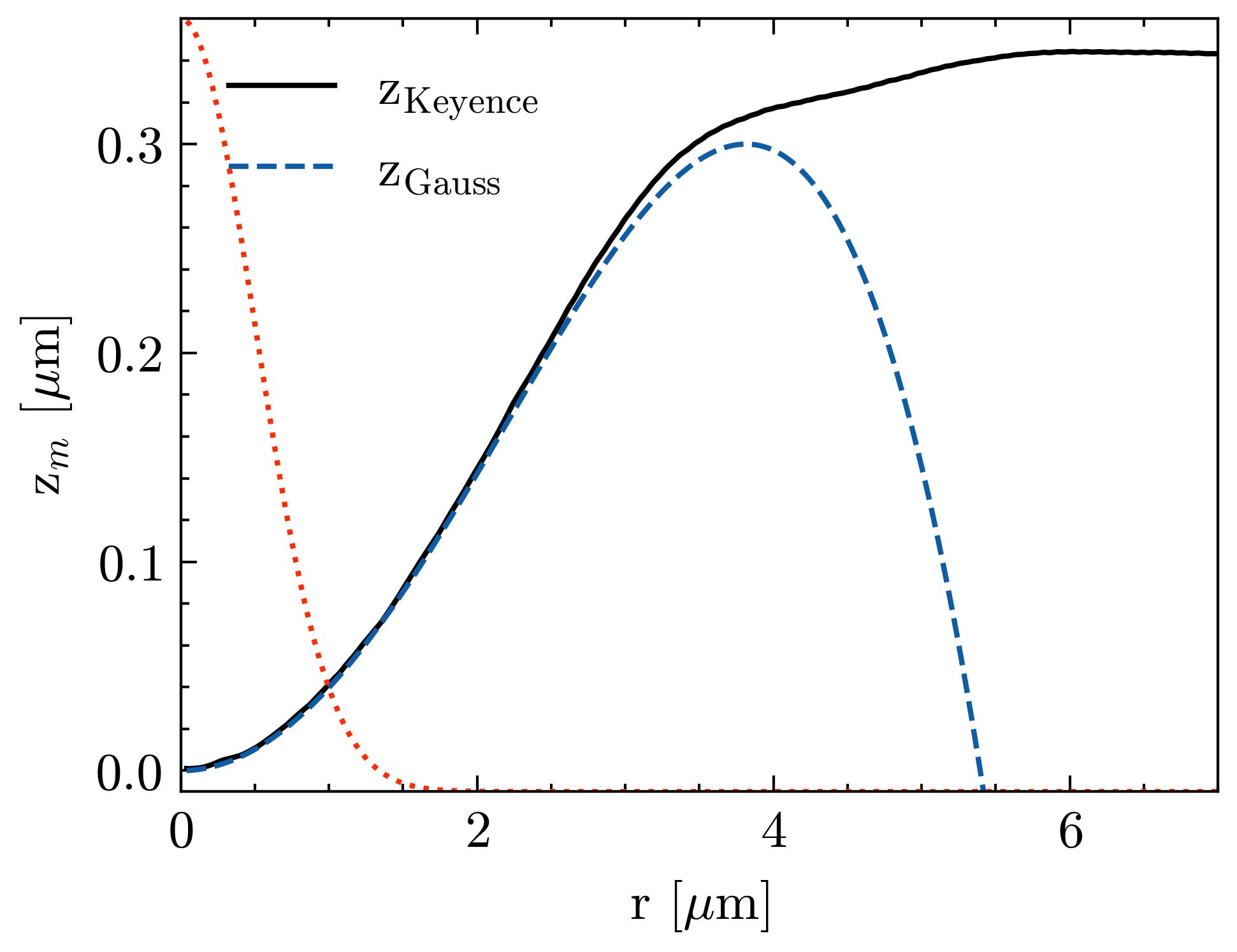}
    \caption{Radial height profile of mirror. The black solid curve is measured with a Keyence interference microscope. The blue dashed curve presents the Taylor expansion of the fitted mirror shape for $R_m = 12.1~\mu$m and $h = 0.60~\mu$m as `felt' by the fundamental mode according to Eq. \eqref{eq:z_rms}. The red dotted curve visualizes the weight $\alpha(r) \propto I(r)$, for $L_{\rm cav} = 1.6\ \mu \text{m} \approx 5\ \lambda/2$.
    }
    \label{fig:MirrorRadialProfile}
\end{figure}

We have also inspected the mirror profile with a Keyence interference microscope. The depth of the mirror is $h = 0.35(1)\ \mu$m, which is close to the mirror depth of $h=0.30(1)$ determined from comparison with the planar modes (see Sec. \ref{subsec:MirrorDepth}).
A fit to Eq. (\ref{eq:z}) yields an estimated $\ea = 0.02(1)$ and the rotation-averaged height profile shown in Fig. \ref{fig:MirrorRadialProfile}. 
The mirror shape is close to a Gaussian, but is more flat in the center and a bit steeper along its sides. 
In short, its shape somewhat resembles a bathtub. 


When fitting the measured height profile, 
the obtained estimates depend on the fit area. 
In general, we fit the mirror shape by minimizing
\begin{eqnarray} \label{eq:z_rms}
    \langle \Delta z^2 \rangle \propto \int (z_{\rm Keyence}-z_{\rm Gauss})^2 \alpha(r) dxdy\,, \ \ 
\end{eqnarray}
for $z_{\rm Gauss}(r) = \frac{r^2}{2R_m} -  \frac{r^4}{8 h_4 R_m^2}$, where $\alpha(r)$ is a weight factor.
We typically fit the mirror shape `felt by a mode' by choosing $\alpha(r) \propto I(r)$, like in Eq. \eqref{eq:delta-z^2}.
For the $N=0$ mode we use $\alpha(r) \propto e^{-r^2/\gamma^2}$ for the value of $\gamma$ expected at $L_{\rm cav} = 1.6\ \mu \text{m} \approx 5\ \lambda/2$, as presented by the red curve in Fig. \ref{fig:MirrorRadialProfile}. 
This fit yields $R_m = 12.1~\mu$m and $h_4 = 0.60~\mu$m, in agreement with  
our earlier findings in Section \ref{subsec:gouy_phase}. This value of $h_4$ is twice as large as the actual measured mirror depth $h$ because it is only based on a fit `in the bottom of the well'.
Modes in larger cavities will have larger waists, will probe larger parts of the mirror, and will thus yield more accurate estimates of the actual mirror depth $h$, with modified values of $R_m$. 
And higher-order modes will also probe larger and different regions of the mirror. 
This explains why the results for the $N=2$ group, presented in Sec. \ref{subsec:FineStructure_complete}, yielded different estimates of $R_m=18.4(2)\ \mu$m and $h_4=0.324(5)\ \mu$m.

The above discussion shows the limitation of a fourth-order Taylor expansion of the mirror shape.
The semi-analytic treatment of associated spherical aberration presented in Sec. \ref{sec:mirror-shape} yield convenient equations for the expected resonances and a good qualitative agreement. 
But quantitative interpretation of the fit parameters $R_m$ and $h_4$ is only straightforward if the mirror shape is really Gaussian and if the cavity modes are compact enough.  
The fit parameter $h_4$ is the most sensitive and should always be interpreted as a convenient parameter to quantify the $r^4$-contribution to the mirror shape, instead of the actual mirror depth $h$. 
The spherical aberration also results in a difference between the mirror radius $R_m$ and the effective radius $R_{\rm eff}$ of the mirror area probed by the mode. 
Nonparaxial effects are bound to scale with $1/R_{\rm eff}$ instead of $1/R_m$, but the difference is often small.

\section{Concluding summary}
\label{sec:summary}

In this paper, we have presented measurements on the optical resonances in tunable plano-concave microcavities.
We have analyzed the observed transmission and reflection spectra as well as the vector-field profiles of several resonant modes.
We have compared these results with theoretical predictions that were presented in Sec. \ref{sec:theory} and revolved around the competition between three mode-shaping effects: (i) mirror anisotropy, (ii) a non-spherical correction due to the Gaussian mirror shape, and (iii) various nonparaxial effects, including spin-orbit coupling. 
This theory contains extensions of the existing theory \cite{VanExter2022} that were needed for the analysis and mode characterization presented. 

The experiments presented in Secs. \ref{sec:spectra} and \ref{sec:profiles} show that all three mode-shaping effects are present in our cavities, although with varying strengths depending on cavity lengths. 
We have demonstrated that the theoretical framework, which models the mirror shape with three parameters, can accurately model the fine structure and predict the vector profiles of the eigen modes.
The analysis has been presented in a step-by-step fashion to provide a recipe for experimentalists to characterize their cavities.
This recipe starts by analyzing the mode spectrum and associated Gouy phases to estimate the longitudinal mode numbers $q$ and mirror radius of curvature $R_m$, and observe potential aspheric effects. 
It then estimates the mirror depth $h$ by comparing the plano-concave cavity spectrum to that of the planar cavity.
After that, it analyzed the $N=1$ transverse mode splitting to estimate the mirror anisotropy in the presence of nonparaxial effects.  
The analysis of high-order ($N \geq 2$) resonances, both spectrally and spatially, was used to improve on these earlier estimates and study the competition between multiple mode shaping effect. 
Notably, the nonparaxial effects arising from spin-orbit coupling are prominently visible in the polarization-resolved profiles of the eigen modes.

The measurements presented in this paper have also been done on other micro cavities with similar micro mirrors.
The results were comparable, but the mirror shapes were obviously different. 
The mirror radii that we obtained were $R_m \approx 15-30~\mu$m, their anisotropy was $\ea \approx 0.02-0.03$, and their mirror depths $h \approx 0.2-0.3~\mu$m.
The spread in these values is modest, basically because these mirrors were produced under nominally identical conditions. 
Still, their precise characterization is important for the selection of the best mirrors, where ``best" typically means the mirror with the lowest losses and hence with the largest depths and the least anisotropy, although different experiments might have different requirements. 

\section{Acknowledgements}
\label{sec:acknowledgements}

This publication is part of the project ``Optical microcavities and 2D quantum emitters" with file number NGF.1623.23.015 of the research programme NGF - Quantum Delta NL Quantum Technologie 2023 which is (partly) financed by the Dutch Research Council (NWO).
The work in Basel received funding from the European Union Horizon 2020 research and innovation program under the Marie Skłodowska-Curie Grant Agreement No. 847471 (Quantum Science and Technologies at the European Campus, QUSTEC) and Grant Agreement No. 861097 (Initial Training Network, QUDOT-TECH), and from the Swiss Nanoscience Institute (SNI).

\bibliographystyle{apsrev4-1}
\bibliography{vanExter}

\appendix
\counterwithin{figure}{section}
\counterwithin{equation}{section}

\section{Calibration of transmission spectra}
\label{sec:appendix-Calibration}

This appendix describes how transmission data can be transformed into $T(L)$ spectra, using the observed transmission peaks as calibration markers for the real cavity lengths.
The expected longitudinal mode spacing
\begin{eqnarray}
    \label{eq:L-long}
    \Delta L_{\text{long}}(q, N) \equiv L(q+1,N) - L(q,N) \nonumber\\ \approx \frac{\lambda}{2} \left( 1 + (N+1)\frac{\lambda}{2\pi} \frac{\partial \chi_0(L)}{\partial L} \right)\,,
\end{eqnarray}
is slightly larger than $\lambda/2$ because $\chi_0(L)$ increases with $L$, but the correction is typically only order 1\% for a typical beam divergence $\theta_0 = \lambda/\pi w_0 = 0.2$. 
We use this mode spacing as a ruler to correct for the nonlinearity of the piezo scan, which occurs mainly at the start of the scan and is caused by hysteresis. 
We then use the expected transverse-mode spacing 
\begin{eqnarray}
\label{eq:L-trans}
    \Delta L_{\text{trans}}(q, N) \equiv L(q,N+1) - L(q,N) \nonumber\\ \approx \frac{\chi_0(L(q,N))}{\pi}\frac{\lambda}{2} \left( 1 + (N+2)\frac{\lambda}{2\pi} \frac{\partial \chi_0(L)}{\partial L} \right) \,.
\end{eqnarray}
as a new division on this ruler.
Division of Eq. (\ref{eq:L-trans}) by Eq. (\ref{eq:L-long}) yields
\begin{equation}
\label{eq:definition-Gouy}
    \frac{\chi_0(L(q,N))}{\pi} = \frac{\Delta L_{\text{trans}}(q,N)}{\Delta L_{\text{long}}(q,N+1)} \,,
\end{equation}
but we generally use $\Delta L_{\text{long}}(q,N)$ in the denominator because it is convenient and the difference is typically too small to notice. 

Equation (\ref{eq:definition-Gouy}) allows one to determine the absolute cavity length $L(q,N)$, and the associated $q$, and the radius of curvature $R_m$.
We start by guessing the offset $\tilde{q}$ in the relation $q = \tilde{q} + \Delta q_{\rm exp}$ and $R_m$, based on prior knowledge or earlier iterations.
This guess yields absolute cavity lengths $L_{\rm guess}(\tilde{q}+\Delta q_{\rm exp},N)$, derived from Eq. (\ref{eq:concave-resonance}), and a relation between $\chi_0$ and $q_{\rm exp}$ of the form
\begin{equation}
\label{eq:sin2}
    \sin^2[\chi_0(L_{\rm guess})] \approx \frac{L_{\rm guess}  + \Delta \tilde{q} \, \lambda/2 + L_{\rm pen}}{R_m} \,, 
\end{equation}
where $\Delta \tilde{q} = \tilde{q} - \tilde{q}_{\rm guess}$ is the difference between the actual offset $\tilde{q}$ and the guessed offset $\tilde{q}_{\rm guess}$.
A plot of $\sin^2[\chi_0(L_{\rm guess})]$ versus $L_{\rm guess}$ and its comparison with Eq. (\ref{eq:sin2}) now yields better estimates of $R_m$ (slope) and the combination $\Delta \tilde{q} \,\lambda/2 + L_{\rm pen}$ (axis crossing). 
These estimates can be used iteratively and should quickly converge to reliable estimates of $R_m$, $\tilde{q}$ and $L_{\rm pen}$.

The combined effective penetration depth $ L_{\rm pen} = (L_{D 1}+L_{D 2}) - (L_{\varphi 1} + L_{\varphi 2})$ introduced in the main text might need some explanation. 
For practical cavities with Distributed Bragg Reflectors (DBRs), the cavity lengths $L$ change into effective cavity lengths that include the optical penetration into the DBRs.
The resonant length in Eq. (\ref{eq:concave-resonance}) now changes to $L(q,N) = L + L_{\varphi 1} + L_{\varphi 2}$, where $L$ is the on-axis distance between the DBRs and the $L_\varphi$'s are their phase penetration depths \cite{Koks2021}.
However, the cavity length $L$ in the equation for the fundamental Gouy phase $\chi_0(L) = \arcsin{\sqrt{L/R_m}}$ now changes from $L$ to $\approx L+L_{D 1}+L_{D 2}$, where the $L_D$'s are modal penetration depths \cite{Koks2021}. 
The approximation is valid for $L \ll R_m$ because the $L_D$ contribution of the curved mirror decreases for larger $L$.
The combined effect of these penetration depths yields 
\begin{equation}
    \sin^2{[\chi_0(L)]} = \frac{L(q,N)+L_{\rm pen}}{R_m} \,, 
\end{equation}
which is Eq. (\ref{eq:Gouy2}) in the main text. 

One might be tempted to use the textbook equation $L_I = \lambda/4(n_H-n_L)$ for the penetration depth into each DBR, but this equation only describes the penetration of the optical intensity.
This penetration depth, for instance, does not affect the distance between the field nodes $L(q,N)$ in Eq. (\ref{eq:concave-resonance}), which instead depend on the frequency detuning relative to the center of the DBR stopband.
And on whether the DBR starts with the low(L) or the high(H) index material, where matched L-DBRs have a phase penetration depth $L_\varphi = \lambda/4$ because they have an anti-node at their surface.
The modal penetration depth in the center of the stopband is approximately equal to $L_I$ in L-DBRs, but is actually \cite{Koks2021,Koks2024} 
\begin{equation}
\label{eq:LD}
    L_{D,L}= (\frac{n_H}{n_L}+\frac{n_L}{n_H}) \frac{\lambda}{8(n_H-n_L)}
\end{equation}
The modal penetration in H-DBRs is a factor $n_H n_L$ smaller at $L_{D, H} = L_{D,L} / n_H n_L$. 

Our cavity comprises one H-DBR (D2) and one modified L-DBR (D1).
Substitution of $n_H = 2.12$ and $n_L = 1.45$ in Eq. (\ref{eq:LD}) yields $L_{D,H} = 0.26 \,\lambda/2$ for our H-DBR and $L_{D,L} = 0.80 \,\lambda/2$ for a perfect L-DBR.
But we use a modified L-DBR, which starts with a $0.8 \times \lambda/4n_L$ thick layer of L medium and has a stopband centered at 610 nm instead of 633 nm.
For this mirror we calculate $L_{D1} \approx 0.40 \,\lambda/2$ at $\lambda = 633$~nm, making $L_{D1} + L_{D2} \approx 0.66 \,\lambda/2$.
The phase penetration depth is zero for our matched H-DBR and would be $\lambda/4 = 0.50 \,\lambda/2$ for a matched L-DBR.
But it is $\approx 0.28 \,\lambda/2$ for our modified L-DBR. 
The combination of these numbers yields $L_{\rm pen} \approx (0.66 -0.28) \,\lambda/2 = 0.38 \,\lambda/2$.

\section{Wavefront matching $\mathbf{\Rightarrow R_{\rm eff}}$}
\label{sec:appendix-Reff}

This appendix describes how the 
aspheric $r^4$-contribution to the mirror shape can be partially compensated for by modifying the beam waist $w_0$. 
More specifically, it uses the (intensity-weighted) shape mismatch between the modal wavefront and the curved mirror to calculate the modified waist of the best-matched mode and its effective radius of curvature $R_{\rm eff}$ and estimated loss upon reflection.
We consider a mirror surface $z_{\rm mirror}=L-z_m$ with 
\begin{equation}
    z_m(r) = \frac{r^2}{2R_m} - \tilde{c} \frac{r^4}{8R_m^3} \,,
\end{equation}
where $\tilde{c} = 0$ for a parabola, $\tilde{c}=-1$ for a sphere, and $\tilde c = R_m/h \gg 1$ for a Gaussian mirror; see Eq. (\ref{eq:z}).
We consider an LG-mode mode with waist $w_0$ at the flat mirror, and width $w(L)$ and radius of curvature $R(L)$ at the curved mirror.
The paraxial wavefront of this mode at the curved mirror
has an approximate paraxial shape 
\begin{equation}
    z_{\rm wave} \approx L_{\rm par} - \frac{r^2}{2R} \,,
\end{equation}
where
\begin{equation}
    L_{\rm par} = q \frac{\lambda}{2}+(N+1)\arcsin{\sqrt{\frac{L}{R}}} \,,
\end{equation}
$R = z + z_0^2/z$ and $z_0^2 = L(R_m-L)$.
Nonparaxial effects, in principle, modify this wavefront into the more general form $z_{\rm wave} \approx  a_{\rm wave} - b_{\rm wave}r^2 - c_{\rm wave}r^4$. 
However, in this appendix we only want to analyze the effect of the typically much stronger $r^4$-term in the mirror shape and, hence, will only use the paraxial form. 
We introduce $s = (r/\gamma)^2$, with $\gamma = \gamma(z) = w(z)/\sqrt{2}$, as integration parameter and write the intensity-weighted mismatch as
\begin{eqnarray}
\label{eq:delta-z^2}
    \langle \Delta z^2 \rangle & = & \int (z_{\rm wave}-z_{\rm mirror})^2 I(s) ds\,, \\
\label{eq:delta-z^2-2}
    & = & \int (a + bs + cs^2)^2 I(s) ds\,,
\end{eqnarray}
where $a = L-L_{\rm par}$, $b = \gamma^2(1/2R-1/2R_m)$, and $c=\tilde{c} \gamma^4/(8R_m^3)$. 
The intensity profile \cite{VanExter2022}
\begin{equation}
    I_{p\ell}(s) = \frac{p!}{(p+\ell)!} s^\ell [{\cal L}^\ell_p(s)]^2 e^{-s} \,.
\end{equation}

For a given mirror shape $\tilde{c}$, we wonder which choice of $w_0$ and $L$ will minimize $\langle \Delta z^2 \rangle$.
For the fundamental mode, $I(s) = \exp(-s)$, a straightforward integration of Eq. (\ref{eq:delta-z^2-2}) yields a quadratic equation in $a$, $b$ and $c$ with a minimum at $b = -4c$ and $a = 2c$.
The first equation, $b = -4c$, yields
\begin{equation}
    \frac{1}{R_{\rm eff}} \approx \frac{1}{R_m} \left( 1 - \frac{1}{kh} \sqrt{\frac{L}{R_m-L}} \right) \,,
\end{equation}
for $L \ll R_m$.
This is Eq. (\ref{eq:Reff}) in the main text. 
The second equation, $a = 2c$, yields
\begin{equation}
\label{eq:L-asp}
    \Delta L_{asp} = - \tilde{c} \frac{\gamma^4}{4R_m^3} \,.
\end{equation}
This result is equal to the expected $\Delta L_j = \langle z_m \rangle$ from Eq. (\ref{eq:asphere}) in the main text, because $\tilde{c} = R_m/h$ and 
$f(p,\ell) = 2$ for $N=0$.

The optimum choice of $w_0$ and $L$, or equivalently $a$ and $b$, does not reduce $\langle \Delta z^2 \rangle$ to zero, but to
\begin{equation}
\label{eq:residual}
    \langle \Delta z^2 \rangle_{\rm min} = (2c)^2 = \Delta L_{\rm asp}^2 \,.
\end{equation}
This mismatch results in a round trip intensity loss \cite{Bennett1992}
\begin{equation}
\label{eq:A-scatter}
    {\rm A}_{\rm scatter} = \int [2k\Delta z(x,y)]^2 I(x,y) dxdy\,.
\end{equation}
because the modal reflection amplitude $\langle \exp(-i\Delta \varphi) \rangle = \exp(-\frac{1}{2} \langle \Delta \varphi \rangle)$, with $\Delta \varphi = \varphi(x,y) = 2 k \Delta z(x,y)$, where $\langle \rangle$ indicates averaging over the modal intensity profile $I(x,y)$. 
Substitution of Eq. (\ref{eq:residual}) in Eq. (\ref{eq:A-scatter}) yields 
\begin{equation}
    {\rm A}_{\rm scatter} \approx  \left( \frac{L}{4kh(R_m-L)} \right)^2 \,. 
\end{equation}
This is Eq. (\ref{eq:loss1}) in the main text.


We can generalize the above result to any $(p,\ell)$ mode by using the orthogonality of the generalized Laguerre-Gauss polynomials and the relation
\begin{equation}
    s {\cal L}^\ell_p = -(p+\ell) {\cal L}^\ell_{p-1} + (2p+1+\ell) {\cal L}^\ell_p - (p+1) {\cal L}^\ell_{p+1} \,.
\end{equation}
The calculation only requires careful bookkeeping and minimization of a quadratic polynomial in $a$ and $b$. 
The generalized result can be expressed as
\begin{equation}
    \frac{1}{R_{\rm eff}} - \frac{1}{R_m} \approx - \tilde{c} \frac{\gamma^2}{R_m^3} C(p,\ell) \,, 
\end{equation}
\begin{equation}
    C(p,\ell) = \frac{p(p+\ell)N +p_+(p_++\ell)(N+2)}{p(p+\ell)+p_+(p_++\ell)} \,, 
\end{equation}
where $p_+ = p+1$ and $N=2p+\ell$. 
We thus find that the difference between the effective curvature $1/R_{\rm eff}$ of higher-order modes relative to the paraxial $1/R_m$ is roughly proportional to $N$, although the final expression is slightly more complicated and also depends on $p$ and $\ell$ individually. 
The generalized result for the modal loss
is too ugly to display. 
It is the ratio of a sixth-order polynomial over a quadratic function in $p$ and $\ell$ and hence it scales roughly with $N^4$. 

We finish this appendix with an alternative approach to calculate the modal loss of the fundamental mode. 
This pragmatic approach is based on the idea that loss occurs when the outer wings of the intensity profile ``don't fit on the curved mirror anymore'', just as loss from an aperture occurs because the aperture truncates part of the beam. 
Suppose that we interpret the fitting criterion, admittedly rather ad hoc, as the criterion that the intensity should fit in a disk with a radius determined by the deflection points of the mirror shape where $d^2 z_m(r)/dr^2 = 0$, being a disk of radius $\sqrt{hR_m}$.
An easy integration then predicts an intensity loss 
\begin{equation}
    A_{\rm deflection} \approx \exp{\left( -hk\sqrt{\frac{R_m-L}{L}} \right)} \,,
\end{equation}
for the fundamental mode.
This is Eq. (\ref{eq:loss2}) in the main text. 
Whether this equation is reasonable remains to be seen.

\section{Polarization tomography}
\label{sec:appendix-Tomography}

This appendix describes the shape birefringence created by spin-orbit coupling on an anisotropic mirror. 
The two linearly polarized $N=0$ modes can lose their frequency degeneracy because the vector correction $\Delta L_{\rm non} = - \Delta L_n = - \lambda^2/(8\pi^2 R_m)$ for $\ell =0$ in Eq. (\ref{eq:nonparaxial}) depends on the mirror radius $R_m$ \emph{in the polarization direction}. 
Using our earlier definition for $\ea$, this makes the vector corrections of the $x/H$ and $y/V$-polarized modes equal to $L_{H,V} \approx - \Ln (1 \pm \ea)$, with average $\Ln$ and splitting $\Delta L_{HV} = L_H - L_V = - 2 \Delta \ea \Ln$. 
Below, we will assume that the two $N=0$ mode have equal (HWHM) width $\delta L$ and denote their average resonance length by $\overline{L}$. 

Even when the two $N=0$ modes cannot be resolved because $\Delta L_{HV} < \delta L$, polarization-resolved measurements can still yield their splitting.
This procedure, which uses polarization tomography, works as follows. 
Suppose that the input field 
has a linear polarization at an angle $\theta_{in}$ with respect to the mirror $x$ axis, such that $\vec{e}_{in} = \cos{\theta_{in}} \vec{e}_x + \sin{\theta_{in}} \vec{e}_y$ and suppose that we observed the transmission behind an output polarizer at an angle $\theta_{out}$. 
The amplitude transmission of the $H$ polarized mode will then be
\begin{equation}
\label{eq:tH}
    \frac{t_H}{t_{max}}= \frac{\cos{\theta_{in}} \cos{\theta_{out}}}{1 - i(L-L_H)/\delta L}  \,.
\end{equation}
The amplitude transmission of the $V$ polarized mode is described by a similar equation with $\sin{\theta_{in}} \sin{\theta_{out}}$ in the numerator and $L_H$ replaced by $L_V$ in the denominator. 
If $\theta_{\rm in} = 0^\circ$ or $90^\circ$, the input will excite only one of the two modes and the transmission $T(L)$ will have a simple Lorentzian shape and scale with scale with $\cos^2\theta_{\rm out}$ or $\sin^2\theta_{\rm out}$.
But a more general input polarization will excite both modes and the projection by the output polarizer can probe their interference. 
More specifically, a Taylor expansion of Eq. (\ref{eq:tH}) and a related equation for $t_V/t_{\rm max}$ for $\Delta L_{HV} \ll \delta L$ produces the combined polarization-projected amplitude spectrum
\begin{eqnarray}
    \frac{t(L)}{t_{max}} & \approx & \frac{\cos{(\theta_{in}-\theta_{out})}}{1 - i(L-\overline L)/\delta L} - \nonumber \\
    & & \frac{i \Delta L_{HV}}{2 \delta L} \, \frac{\cos{(\theta_{in}+\theta_{out})}}{[1 - i(L-\overline L)/\delta L]^2} \,.
\end{eqnarray}

The first term describes the transmission spectrum of the ``parallel-polarization" component, with $\cos{(\theta_{in}-\theta_{out})}$ via Malus law.
The second term describes the transmission spectrum of the ``perpendicular-polarization" component.
For orthogonal input/output polarizations at $\pm 45^\circ$ with respect to the long/short axes of the mirror, the resulting transmission spectrum
\begin{equation}
    \frac{T_\perp(L)}{T_{max}} = \left( \frac{\Delta L_{HV}}{2 \delta L} \right)^2 \frac{1}{[1 + (L-\overline L)^2/\delta L^2]^2} \,,
\end{equation}
has a different shape and is sharper than the ``parallel" Lorentzian spectrum.
A measurement of the ratio of the $T_\perp(L)$ and $T_\parallel(L)$ peaks thus allows one to calculate the splitting $\Delta L_{HV}$, even if the two polarized modes are not spectrally resolved, and compare this with the theoretical prediction
\begin{equation}
    \frac{\Delta L_{HV}}{2 \delta L} = \ea \frac{F\lambda}{2\pi^2 R_m} \,.
\end{equation}
Measurements at different polarization angles are expected to produce asymmetric transmission peaks. 

\end{document}